\documentclass[11pt,a4paper]{article}

\usepackage{jheppub,amsmath,  amssymb,slashed,url,bm,textgreek,upgreek}
\usepackage{graphicx}
\usepackage{epstopdf}
\usepackage{braket}
\UseRawInputEncoding

\def\t{{ \sf t}} 

\def\tt{{\mathfrak t}}

\def\a{{\sf a}}
\def\b{{\sf b}}

\def\dim{{\mathrm{dim}}}

\def\s{{s}}

\def\max{{\mathrm{max}}}

\def\x{{\sf x}}
\def\y{{\sf y}}
\def\mod{{\mathrm{mod}}}

\def\be{\begin{equation}}
\def\ee{\end{equation}}

\def\hat{\widehat}

\def\tilde{\widetilde}

\def\h{\widehat}

\def\vol{{\mathrm{vol}}}

\def\y{{\mathrm y}}

\def\O{{\mathcal O}}

\def\K{{\mathcal K}}

\def\A{{\mathcal A}}
\def\d{{\mathrm d}}

\def\b{\overline}
\def\R{{\mathbb R}}
\def\C{{\mathbb C}}
\def\U{{\mathcal U}}

\def\[{\bigl [}

\def\]{\bigr ]}

\def\Z{{\mathbb Z}}

\def\t{\widetilde }
\def\h{\widehat}

\def\hor{{\mathrm{hor}}}

\def\B{{\mathcal B}}

\def\P{{\mathcal P}}
\def\l{\langle}

\def\H{{\mathcal H}}

\def\vol{{\mathrm{vol}}}

\def\tilde{\widetilde}

\def\bar{\overline}

\font\teneurm=eurm10 \font\seveneurm=eurm7  \font\fiveeurm=eurm5
\newfam\eurmfam
\textfont\eurmfam=\teneurm \scriptfont\eurmfam=\seveneurm
\scriptscriptfont\eurmfam=\fiveeurm

\font\teneusm=eusm10 \font\seveneusm=eusm7 \font\fiveeusm=eusm5
\newfam\eusmfam
\textfont\eusmfam=\teneusm \scriptfont\eusmfam=\seveneusm
\scriptscriptfont\eusmfam=\fiveeusm

\font\tencmmib=cmmib10 \skewchar\tencmmib='177
\font\sevencmmib=cmmib7 \skewchar\sevencmmib='177
\font\fivecmmib=cmmib5 \skewchar\fivecmmib='177
\newfam\cmmibfam
\textfont\cmmibfam=\tencmmib \scriptfont\cmmibfam=\sevencmmib
\scriptscriptfont\cmmibfam=\fivecmmib

\def\rel{{\mathrm{rel}}}
\def\grav{{\mathrm{grav}}}
\def\dS{{\mathrm{dS}}}
\def\SO{{\mathrm{SO}}}
\def\la{\langle}
\def\ra{\rangle}
\def\obs{{\mathrm{obs}}}
\def\matt{{\mathrm{matt}}}
\def\i{{\mathrm i}}
\def\cr{{\mathrm{cr}}}
\def\Tr{{\mathrm{Tr}}}
\def\b{{\sf b}}
\def\TFD{{\mathrm{TFD}}}
\def\gen{{\mathrm{gen}}}
\def\out{{\mathrm{out}}}

\def\hat{\widehat}

\def\Ad{{\mathrm {Ad}}}
\def\f{\varphi}
\def\l{\lambda} 
\def\al{\alpha}

\title{An Algebra of Observables for de Sitter Space}

 \author{Venkatesa Chandrasekaran,$^1$   Roberto Longo,$^2$ Geoff Penington,$^{1,3}$ and  Edward Witten$^1$}
\affiliation{$^1$School of Natural Sciences, Institute for Advanced Study,\\ 1 Einstein Drive, Princeton, NJ 08540 USA}
\affiliation{$^2$Dipartimento di Matematica, Universita di Roma �Tor Vergata�,
Via della Ricerca Scientifica, 1, I-00133 Roma, Italy}
\affiliation{$^3$Center for Theoretical Physics and Department of Physics, University of California,\\  Berkeley, CA 94720 USA}
\abstract{We describe an algebra of observables for a static patch in de Sitter space, with operators gravitationally dressed to the worldline of an observer.
The algebra is a von Neumann algebra
of Type II$_1$.   There is a natural notion of entropy for a state of such an algebra.
There is a maximum entropy state, which corresponds to empty de Sitter space, and the entropy of any semiclassical state of the Type II$_1$
algebras agrees, up to an additive constant independent of the state, with the expected generalized entropy $S_\gen=(A/4G_N)+S_\out$.     An arbitrary additive
constant is present because of the renormalization that is involved in defining entropy for a Type II$_1$ algebra.
}

\begin{document}\maketitle

\section{Introduction}\label{intro} 

\subsection{Background}

Not long after it was understood that an entropy should be associated to the horizon area of a black hole  \cite{Bekenstein,Hawking}, Gibbons and Hawking \cite{GH}
proposed that similarly, the area of a cosmological horizon should be interpreted as an entropy.  Specifically, they considered an observer in  a de Sitter space with
radius of curvature $r_\dS$.   The worldline of the observer is assumed to be timelike.   The region of de Sitter space that is causally accessible to such an observer
is bounded by past and future horizons.   Gibbons and Hawking associated to the horizon of an observer 
an entropy $A/4G_N$  (where $A$ is the area of the observer's horizon and $G_N$ is Newton's constant)
and a temperature $T_\dS=1/2\pi r_\dS =1/\beta_\dS$.   The  temperature, but not the entropy, had been defined earlier by Figari, 
Hoegh-Krohn, and Nappi \cite{FHN}.

A difference between a black hole horizon and a cosmological horizon is that, in a sense, the notion of a cosmological horizon is more observer-dependent.
A black hole horizon in an asymptotically flat (or asymptotically Anti de Sitter) spacetime is defined in terms of the region that is causally accessible to
any observer at infinity.   In a cosmological model such as de Sitter space, different observers can see and can influence different parts of the spacetime, and
experience different horizons.  De Sitter space has a great deal of symmetry, such that the horizon of any observer has the same area.   That is not true
in a more general cosmological model.  

It seems fair to say that although black hole entropy remains highly enigmatic to this day,  the entropy of a cosmological horizon, such as the de Sitter horizon,
 is only more mysterious.
A working hypothesis, expressed in a modern form in \cite{Review}, is that black hole entropy measures the logarithm of the dimension
 of a quantum Hilbert space that is needed
to describe a black hole as seen by an observer who remains outside the horizon.
 What would be the analog of this for cosmological horizons?   A plausible   analog of an observer who remains outside the black hole horizon is an observer in
 de Sitter space
 whose horizon is under discussion.    The region of de Sitter space causally accessible to such an observer has been called a ``static patch.''   
 Thus a possible analog of the working hypothesis about black hole entropy would be to say that the de Sitter entropy measures the logarithm of the dimension
  of the quantum
 Hilbert space that an observer in the static patch can use to account for  inaccessible degrees of freedom beyond the horizon.
 
 Something somewhat along these lines has actually been proposed \cite{BoussoOne,BoussoTwo,Banks,BanksFischler,BanksOne,BanksTwo,BFTwo,SusskindA,Susskind}, but with a subtle difference.    To explain this point,
 let us consider a de Sitter space that instead of being empty, as considered originally by Gibbons and Hawking, contains ordinary particles and fields and perhaps
 even some small  black holes.   In this case, we follow Bekenstein and define a generalized entropy that includes the horizon entropy $A/4G_N$ and also the
 ordinary entropy $S_\out$ of the matter that the observer can see:
 \be\label{sgen}S_\gen=\frac{A}{4 G_N} +S_\out. \ee
 In the case of a black hole in an asymptotically flat spacetime, both terms $A/4G_N$ and $S_\out$ can be arbitrarily large, and it takes an 
 infinite-dimensional Hilbert space to describe the possible states of the world as seen by an outside observer.   In the case of de Sitter space, it has been
 argued that this is not the case:  it is claimed that the maximum possible value for $S_\gen$ is the value that it has for empty de Sitter space \cite{Maede,BoussoOne,BoussoTwo}.    One can certainly
 increase $S_\out$ by considering a state in which the static patch is not empty. The claim is that this has the effect of reducing the horizon area, in such a way that the
 decrease in $A/4G_N$ exceeds the increase in $S_\out$.   
 
 The proposal, then, is that empty de Sitter space maximizes the entropy of any state of the static patch. 
  Here by empty de Sitter space we mean the generalization with gravity included of the natural de Sitter invariant state of quantum fields in de Sitter space  \cite{CT,SS,BD,Mo,Al}, which
 is often called the Bunch-Davies state.   
 Thus in leading order for small $G_N$, 
 the maximum possible entropy of a state of the static patch, including particles and fields and black holes
  it may contain and also degrees of freedom
 that are somehow associated to the cosmological horizon, is $A_\dS/4G_N$, where $A_\dS$ is the horizon area  of empty de Sitter space.   (For the one-loop correction
 to the formula $A_\dS/4G_N$ for the entropy of empty de Sitter space, see \cite{Denef}.)
 
 The existence of  a maximum entropy state of de Sitter space makes possible an interpretation of de Sitter entropy that does not quite
 have an analog for black hole entropy.   The proposal is that a Hilbert space of dimension roughly $\exp(A_\dS/4G_N)$ suffices to describe all possible
 states of the static patch, including any matter and black holes it may contain and also the cosmological horizon \cite{BoussoOne,BoussoTwo,Banks,BanksFischler,BanksOne,BanksTwo,BFTwo,SusskindA,Susskind}. Empty de Sitter space would then be described by the maximally mixed state on this finite-dimensional Hilbert space \cite{DST, SusskindB}.
By contrast, in the case of  a black hole, the Bekenstein-Hawking entropy $A/4G_N$ possibly determines the size of a Hilbert space that describes the black hole as seen
 from outside, but this Hilbert space does not describe particles and fields outside the black hole horizon.

The claim that empty de Sitter space is maximally mixed may sound counterintuitive, since the Bunch-Davies state of quantum fields reduces on the static patch to a thermal ensemble with temperature $T_{\dS}$. The idea is that this thermal distribution comes purely from entropic, rather than energetic, suppression on the full static-patch Hilbert space \cite{BanksTwo, BanksOne,Susskind}. 
As a simple illustration, consider a particle with energy $E$ at rest in the center of the static patch. The presence of such a particle
reduces the  area of the cosmological horizon from $A_\dS$ to 
\be\label{horball}
A_{\mathrm{hor}}/4G = A_{\dS}/4G - \beta_{\dS} E. 
\ee
We will
recall the  derivation of this statement in section \ref{bulk}. If the total number of microstates is $\exp(A_\dS/4G)$, and 
the number of microstates such that the particle has energy $E$ is $\exp(A_{\mathrm{hor}}/4G)$, then assuming that 
each microstate is equiprobable,  the probability to observe a particle of energy $E$ in the center of the static patch will be 
\be\label{norball}
p(E) =  \exp(-\beta_{dS} E).\ee
In this way, one can obtain a thermal distribution purely on entropic grounds.
 
 \subsection{A Type II$_1$ Algebra For The Static Patch}
 
 In the present article, we will make a contribution to the understanding of de Sitter entropy by defining a von Neumann algebra of Type II$_1$
 that, in the limit of small $G_N$, describes the possible observations of an observer in de Sitter space.   We will also reconsider and generalize a previous
 discussion  in which, inspired by considerations involving the large $N$ limit
  in holography \cite{LL,LLtwo},  observations outside a black hole horizon were described by an algebra of Type II$_\infty$ \cite{GCP}.
  
  The entanglement entropy of a local region in quantum field theory is always ultraviolet divergent, as discovered long ago   \cite{Sorkin, BKLS}.
  An abstract explanation of why this happens is that the algebra of observables in a local region in quantum field theory is of Type III \cite{Araki},
  and there is no notion of entropy for a state of an algebra of Type III.   By contrast, for states of an algebra of Type II, it is possible to define an entropy,
  though in physical terms this is a sort of renormalized entropy with a state-independent divergent constant subtracted.   Thus, at least
  for the black hole and de Sitter space, the fact that gravity converts the algebra of observables from being of Type III to being of Type II gives an abstract explanation
  of why the entropy of a region of spacetime is better-defined in the presence of gravity.   
  
  A Type II$_1$ algebra has a state of maximum entropy, as has been explained long ago \cite{Segal}
  and  reconsidered recently \cite{LongoWitten}.  Hence a Type II$_1$ algebra is a candidate for describing the physics of a static patch in de Sitter space,
  with empty de Sitter space corresponding to the state of maximum entropy.     By contrast, there is no upper bound on the entropy of a state of a Type II$_\infty$
  algebra.   Therefore, a Type II$_\infty$ algebra is a  candidate for describing physics outside the black hole horizon.  
  
  In section \ref{algstatic}, we consider the problem of defining, in the limit $G_N\to 0$, an algebra of observables in a static patch of de Sitter space.   
  There is no asymptotic region at infinity to which a local operator can be ``gravitationally dressed,'' so we assume the existence of an observer in the static
  patch and we gravitationally dress operators in the static patch to the worldline of the observer.   We consider a minimal model in which the observer consists
  only of a clock.   The resulting algebra is of Type II$_1$.   The maximum entropy state of the Type II$_1$ algebra is the natural state of empty de Sitter space
  (tensored with a simple state of the observer).     The density matrix of the maximum entropy state is the identity, in keeping with the idea that this
  state is maximally mixed and has a flat entanglement spectrum \cite{DST}.
  
  In section \ref{bulk}, we assume that the de Sitter space is not necessarily empty and show that (up to an additive constant that is independent of the state)
  the entropy of a semiclassical state of the Type II$_1$ algebra agrees with the generalized entropy $S_\gen=A/4G_N+S_\out$.    In this, we closely follow an analysis we
  present elsewhere for the case of the Type II$_\infty$ algebra of a black hole  \cite{CPW}.
  We also make use of a formula of Wall expressing the generalized entropy in terms of relative entropy on the horizon \cite{Wall}.
  
  An observer in a static patch in de Sitter space has access to observables that are localized in that patch, but has no way to know what there is beyond the
  horizon and therefore has no knowledge of a global quantum state of the whole system.   However, with some assumptions about what is beyond the horizon,
  one can construct a global Hilbert space for the whole system, and the Type II$_1$ algebra 
  constructed in section \ref{algstatic} should act naturally on this Hilbert space.     We explore this question in section \ref{hilbertspaces}.   We get a simple answer if
  we assume the existence behind the horizon of a second observer who is completely or almost completely entangled with the observer in the static patch.  
  In the absence of such a second observer, the question is more difficult and appears not to have a simple limit for $G_N\to 0$.   
  
  Finally, in section \ref{bh} we reconsider the Type II$_\infty$ algebra of the black hole.   We define this Type II$_\infty$ algebra in a way that makes
  sense for a black hole in an asymptotically flat spacetime (as opposed to the asymptotically AdS spacetime considered in \cite{GCP}).   We also formulate
  the derivation in a way that shows a close analogy between the definition of the Type II$_\infty$ algebra for the black hole and the Type II$_1$ algebra
  for de Sitter space.   
  
  Appendix \ref{takesaki} is an explanation of the crossed product construction for operator algebras and of 
  Takesaki duality, which underlies the construction of the Type II$_1$ algebra for de Sitter space. 
  In Appendix \ref{coinvariants}, we explain how, in the presence of gravity,
   the symmetries of de Sitter space can be imposed as constraints in the construction of a Hilbert space of quantum states.  Here we follow previous
   analysis \cite{higuchi,marolf}, the only novelty being to present the construction in the BRST framework; it is known in general that this is possible
   \cite{shvedov}.   We also explain why it is more straightforward to impose constraints on operators than on states.  Some facts explained in the appendix
   are important background to this article, though  we explain more than we strictly need.

\subsection{What is a Type II$_1$ Algebra?}\label{whatis}

  In the remainder of this introduction, we provide a short introduction to Type II$_1$ algebras.
  
    A Type II$_1$ algebra is just the natural algebra of observables that acts on a countably infinite set of qubits in a state that is almost maximally mixed.
  In more detail,\footnote{This construction goes back to Murray and von Neumann; for a slightly
 different description, see for example section 3.3 of \cite{Curved}.}
consider a system $A$ consisting of a countably infinite set of qubits that are almost completely 
entangled with an identical system $B$ also consisting of a countably infinite set of qubits.  In the case of a finite number $N$ of qubit pairs, a state in which
the $n^{th}$ qubit of system $A$ is maximally entangled with the $n^{th}$ qubit of system $B$ for all $n$  is
\be\label{wacko}\Psi_{\TFD}=\frac{1}{2^{N/2}}\bigotimes_{n=1}^N \sum_{i=1,2}|i\ra_{n,A}\otimes |i\ra_{n,B}. \ee
We have called this state $\Psi_\TFD$ because it can be viewed as the usual thermofield double state specialized to the case that the Hamiltonian is $H=0$.
In the limit $N\to\infty$, one constructs a Hilbert space $\H_\TFD$ with the property that for any state in this space, almost all of the qubit pairs are almost
completely entangled in the same way as in $\Psi_\TFD$.  Let $\a$ be any operator that acts on only the first $k$ qubits of system $A$ and define
\be\label{zoff} F(\a)=\la\Psi_\TFD |\a|\Psi_\TFD\ra. \ee
   Since $\Psi_\TFD$ has been chosen
so that the qubits of the $A$ system are maximally mixed, the density matrix of the $A$ system is $\rho_A=2^{-N} \,I$, where $I$ is the identity matrix.
So 
\be\label{traceform} F(\a)= \Tr\,\rho_A \a =2^{-N}\Tr\,\a,\ee  
and therefore, if  $\a,\a'$ are two operators that both act only on the first $N$ qubits of system $A$, then
\be\label{roff} F(\a\a')=F(\a'\a)=2^{-N}\,\Tr\,\a\a'. \ee
Some other important properties of $F(\a)$ are that it is normalized to
\be\label{woffo} F(1)=1 \ee
and that it is positive, in the sense that
\be\label{zoffo}F(\a^\dagger\a)>0, ~~\a\not=0.\ee

The function $F(\a)$ is defined as soon as one includes all the qubits that $\a$ acts on in the definition of $\Psi_\TFD$,
and it is unaffected by including additional qubits, so it is well-defined in the limit of an infinite system. 
Eqn. (\ref{roff}) says that the  function $F(\a)$ has the algebraic property of a trace, and it is in fact convenient to denote it as one:
\be\label{omoff} F(\a)=\Tr\,\a. \ee Clearly then
\be\label{poff} \Tr\,1=1\ee
and
\be\label{woff}\Tr\,\a^\dagger\a>0,~~\a\not=0.\ee
So far we have defined the trace for the algebra $\A_0$ of all operators that act on only finitely many qubits of system $A$.   However, the definition
can be extended to a larger algebra $\A$ that consists of all operators that can be approximated sufficiently well by operators that act on only finitely
many qubits of system $A$.  (Technically, $\A$ contains operators that are weak limits of sequences of operators in $\A_0$.)   The algebra $\A$ is known
as a von Neumann algebra of Type II$_1$.    The center of $\A$ consists only of complex scalars.   In von Neumann algebra language, this means that
$\A$ is a ``factor,'' somewhat analogous to a simple Lie group.

We are quite familiar with a more elementary example of an infinite algebra that has a trace, namely the algebra $\B$ of all (bounded) operators on a Hilbert space
$\H$ of countably infinite dimension.   This algebra has a trace, but it is not defined for all elements, only for those that are ``trace class''; in particular, in the algebra
$\B$, the trace of the identity operator is $+\infty$.   By contrast, a Type II$_1$ algebra has a trace that is defined for all elements, and which can be normalized
so that $\Tr\,1=1$.   We will see that it is natural to define a Type II$_1$ algebra associated to de Sitter space.   The tensor product of the two algebras
$\A$ and $\B$ that we have described so far is an algebra ${\mathcal C}=\A\otimes \B$ that turns out to be a von Neumann algebra of Type II$_\infty$. ${\mathcal C}$ is a factor, since $\A$ and $\B$ are, and it has a trace, since $\A$ and $\B$ do, but the trace is not defined for all elements of ${\mathcal C}$,
since that is the case for $\B$.    In an asymptotic
expansion near $N=\infty$, the algebra of observables outside a black hole horizon is of Type II$_\infty$ \cite{GCP}.  This was found by incorporating
some $1/N$ corrections in a construction of emergent Type III algebras in holographic duality \cite{LL,LLtwo}.    Our main result in the present
article is an analogous statement involving the static patch of de Sitter space and an algebra of Type II$_1$.

Though not the main focus of the present article,
Type III von Neumann algebras can be constructed in a similar way, starting with the thermofield double state for a nonzero Hamiltonian.  The Hamiltonian
can be taken to be a simple sum of single qubit Hamiltonians, $H=\sum_i H_i$, where $H_i$ acts only on the $i^{th}$ qubit.  In the thermodynamic limit, another novel
 algebra, now of Type III,  can be defined just as before.   The main difference is that this algebra does not have a trace.  See for example \cite{Curved} for a fuller
 explanation.

A Type II or Type III von Neumann algebra does not have an irreducible representation in a Hilbert space.   Whenever such an algebra acts on a Hilbert space $\H$,
it commutes with another algebra $\A'$ that is also of the same type, Type II or Type III.   For example, in the Murray-von Neumann construction of the Type II$_1$
factor $\A$ that we have just described, $\A$ obviously commutes with an isomorphic algebra $\t\A$ that acts on system $B$.   If $\Lambda$ is a projection operator
in $\t\A$, then $\Lambda\H$ is a subspace of $\H$ on which $\A$ acts.   This motivates the question of how ``small'' $\Lambda$ can be and how much we can shrink
$\H$ while still getting a Hilbert space on which $\A$ acts.  This question was addressed by Murray and von Neumann.   Projection operators $\Lambda$
in a Type II$_1$ algebra $\A$ (or $\t\A$) 
are classified, up to unitary equivalence in the algebra, by their trace, and any value of the trace between 0 and 1 is possible.   (The values 0 and 1 occur
only for $\Lambda=0$ and $\Lambda=1$.)   In the construction of $\A$ from the infinite system of entangled qubits, for any positive integers $k$ and $n$, 
the projection operator
onto a $k$-dimensional subspace of the $2^n$-dimensional Hilbert space of the first $n$ qubits of system $A$ gives, in the large $N$ limit, 
a projection operator  $\Lambda\in \A$ with $\Tr\,\Lambda=k/2^n$.   By taking limits, one can define a projection operator $\Lambda\in\A$ with any trace between 0 and 1.
(In section \ref{second}, we will find another construction of a projection operator in $\A$ with any desired value of the trace between 0 and 1.)   

Murray and von Neumann showed that the Hilbert space representations of the algebra $\A$ are classified by a parameter $d$, called the continuous dimension,
that can be any positive real number or $\infty$.   The Hilbert space $\H_\TFD$ is defined to have $d=1$, and if $\Lambda\in \t\A$ is a projection operator
with $\Tr\,\Lambda=f$, then $\Lambda\H$ has $d=f$.   Finally, $d$ is additive in direct sums, so for example the direct sum of $k$ copies of $\H_\TFD$ has
$d=k$.   By taking $d$ very small, we can make a ``small'' representation of $\A$, but there is no irreducible representation; any representation can always
be reduced further.   The ``trace'' function $\Tr\,\a$ that we defined earlier has the algebraic properties of a trace, but it is not actually the trace in any representation
of $\A$; it is more like a trace with an infinite factor $\left.2^{N}\right|_{N\to\infty}$ removed.

Since an algebra of Type II$_1$ does not have an irreducible Hilbert space 
representation, there is no notion of a pure state of such an algebra.   However, there is a notion of
density matrices and entropies.   This results from the fact that the bilinear form $\Tr\,\a\a'$ on a pair of elements $\a,\a'\in\A$ is nondegenerate (the nondegeneracy
is a consequence of the fact that $\Tr\,\a^\dagger\a>0$ for $\a\not=0$).    So any linear function $F(\a)$ on the algebra is of the form $\Tr\,\a\a'$ for a unique\footnote{More precisely, in general $\a'$ may be an unbounded operator 
 ``affiliated'' to $\A$, meaning that bounded functions of $\a'$ are in $\A$. We will usually omit this qualification.}  $\a'\in\A$.

For example, if $\A$ acts on a Hilbert space $\H$ and $\Psi$ is a vector in $\H$, the function $F(\a)=\la\Psi|\a|\Psi\ra$ is a linear function on $\A$, so there is
a unique $\rho\in\A$ such that
\be\label{dono} \la\Psi|\a|\Psi\ra =\Tr\,\a\rho. \ee   Since $\la\Psi|\a^\dagger\a|\Psi\ra\geq 0$ for all $\a\in\A$, we have $\Tr\,\a^\dagger\a\rho\geq 0$ for all $\a$.
This is described by saying that $\rho$ is positive.   Assuming that the state $\Psi$ is normalized to $\la\Psi|\Psi\ra=1$, we have $\Tr\,\rho=1$.
So $\rho$ is a positive element of $\A$ of trace 1.  
By analogy with the standard terminology in ordinary quantum mechanics, such an element of $\A$ is called a density matrix, and we
define $\rho$ to be the density matrix  of the state $\Psi$, for observations in $\A$.   As a simple example, consider the state $\Psi_\TFD$.   Since by
definition $\la\Psi_\TFD|\a|\Psi_\TFD\ra=\Tr\,\a$, the density matrix of $\Psi_\TFD$ is $\rho=1$.

Conversely, suppose that we are given a density matrix $\rho\in\A$.   In ordinary quantum mechanics, every density matrix has a (highly non-unique) ``purification.''
This statement has an analog for an algebra $\A$ of Type II$_1$.   A purification of a density matrix 
$\rho\in\A$ is a Hilbert space $\H$ with an action of $\A$ and a state $\Psi\in\H$ such that
$\Tr\,\a\rho=\la\Psi|\a|\Psi\ra$ for all $\a\in \A$.    The GNS construction of operator algebra theory 
 (see for example section 3.1 of \cite{Curved})
gives a simple construction of a purification of any state of a Type II$_1$ algebra,
generalizing a standard construction in ordinary quantum mechanics in which a purification is constructed by doubling the Hilbert space.

Once one has density matrices, one can also define entropies; for example, the von Neumann entropy is defined as usual as
\be\label{wono}S(\rho)=-\Tr\,\rho\log\rho.\ee
R\'{e}nyi entropies are defined similarly:
\be\label{renyi}S_\alpha(\rho)=\frac{1}{1-\alpha}\log \Tr\,\rho^\alpha.\ee  These entropies have slightly unusual properties that we will discuss first informally
and then in a more formal way.

 The maximally mixed state of $N$
qubits has von Neumann entropy $N$.    So the state $\Psi_\TFD$
on $N$ qubits, reduced to system $A$, has entropy $N$, and in the $N\to\infty$ limit its entropy diverges.   A state obtained from $\Psi_\TFD$ by
disentangling, say, $k$  qubits has entropy $N-k$, less than $N$ but also divergent for $N\to\infty$.    To define entropies for a state in $\Psi_\TFD$,
reduced to system $A$, we subtract $N$ before taking the large $N$ limit.    With this definition, the maximally mixed state of the infinite system
of qubits has entropy 0 and other states have negative entropy.   This is the usual definition for states of an algebra of Type II$_1$.  In short,
entropy in a Type II$_1$ algebra is a renormalized entropy.   There is a state of maximum entropy, whose entropy is defined to be 0, and
other states have negative entropy.

Now let us see these properties from the definition\footnote{For previous treatments, see \cite{Segal,LongoWitten}.} $S(\rho)=-\Tr\,\rho\log\rho$.   First of all, the state of maximum entropy is $\Psi_\TFD$, corresponding to $\rho=1$.
For $\rho=1$, we do indeed have $S(\rho)=0$.  To show that any other density matrix $\sigma$ has $S(\sigma)<0$, one way is as follows.  In general
   let $\rho$ and $\sigma\not=\rho$ be two density matrices.  As in ordinary quantum mechanics, one can interpolate between $\sigma$ and $\rho$ by the one-parameter
   family of density matrices $\rho_t=(1-t)\rho+t\sigma$, $0\leq t\leq 1$.  Let $f(t) =S(\rho_t)$.    The same calculation as in ordinary quantum mechanics
   shows that $f''(t)\leq 0$ for $0\leq t\leq 1$.   In other words, the entropy is concave.   
   If $\rho=1$, one finds $f(0)=f'(0)=0$, $f''(0)= -\Tr\,(1-\sigma)^2<0$. Therefore $f(t)< 0$ for $0< t\leq 1$.
   In particular $S(\sigma)=f(1)<0$.    
   
   As an example that will be relevant in section \ref{algstatic}, let $\Lambda$ be a nonzero projection operator in $\t\A$.  As explained earlier, its
   trace is a positive number that is no greater than 1 (and equal to 1 only if $\Lambda=1$).  Therefore $\rho_\Lambda=\Lambda/\Tr\,\Lambda$ is
   a density matrix.   To evaluate $S(\rho_\Lambda)$, we note
    that the operator $\rho_\Lambda$ has the two eigenvalues 0 and $1/\Tr\,\Lambda$, so $\rho_\Lambda \log \rho_\Lambda=\rho_\Lambda \log(1/\Tr\Lambda)$
    and hence 
    $S(\rho_\Lambda)=-\Tr\,\rho_\Lambda\log\rho_\Lambda = -\log(1/\Tr\,\Lambda)<0$.
    
    The algebra of operators in a local region in ordinary quantum field theory is a von Neumann algebra of Type III, a statement that is more or less equivalent
    to the statement that neither pure states nor density matrices nor entropies can be defined for such an algebra.   The main observation of the present
    article is that, provided an observer is included in the description of the static patch, gravity converts the Type III algebra of the static patch in de Sitter
    space to an algebra of Type II$_1$.

\section{The Algebra Of The Static Patch}\label{algstatic}

\subsection{The Static Patch}\label{static}

What can be seen by an observer in de Sitter space?   We will discuss this question first in the context of ordinary quantum field theory in a fixed
de Sitter background, and then in the context of semiclassical quantization of gravity.

 \begin{figure}
 \begin{center}
   \includegraphics[width=2.5in]{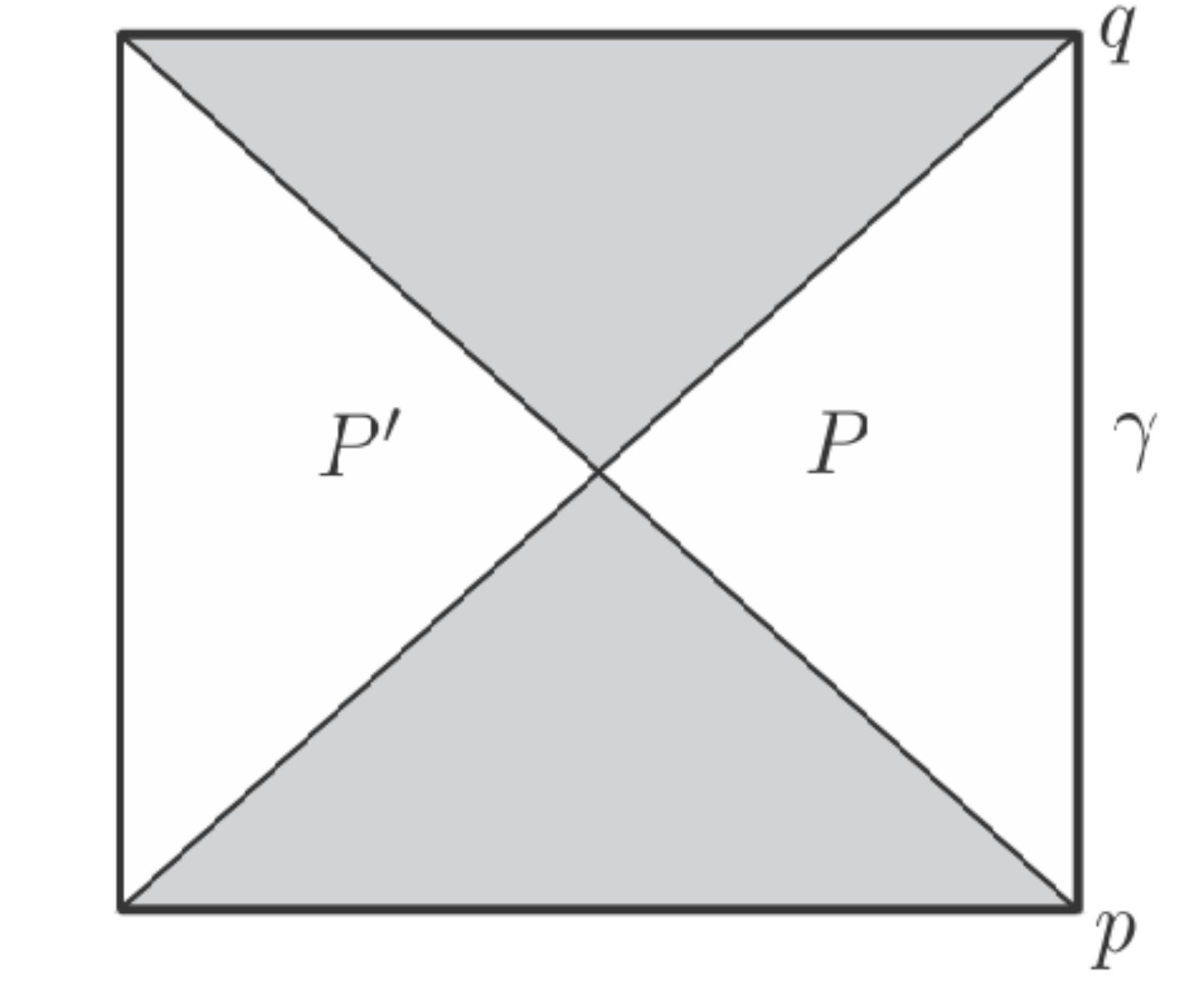}
 \end{center}
\caption{\small The Penrose diagram of de Sitter space.   An observer enters at the point $p$ at past infinity and exits at the point $q$ at future infinity.
Coordinates can be chosen to put these points at corners of the diagram, as shown.
The observer's worldline is a timelike curve $\gamma$ from $p$ to $q$, such as  the geodesic from $p$ to $q$, shown here as
the vertical line on the right boundary of the diagram.     $P$ is the static patch of this observer, and $P'$ is the complementary
static patch, spacelike separated from $P$.   The shaded regions are behind past and future horizons for an observer in either $P$ or $P'$.  \label{AAA}}
\end{figure} 

Imagine an observer who enters a $D$-dimensional  de Sitter space $X$ at a point $p$ in past infinity and exits in the far future at a point $q$ in future infinity (fig. \ref{AAA}).  From $p$ to $q$, the observer travels on some worldline $\gamma$.   The part of de Sitter
space that is causally accessible to the observer -- the part that the observer can influence and can also see, or in other words the intersection of the past
and future of $\gamma$ -- is bounded by past and future horizons.   The causally accessible region, which we will call $P$, is known as the static patch of
the observer.  
$P$ depends only on the points $p$ and $q$ and not
on the path $\gamma$.   The region of $X$ that is spacelike separated from $P$, so that the observer can neither see nor influence it, is a complementary static patch $P'$.  

 Consider first an ordinary quantum field theory
on $X$,  without gravity.   Such a theory has a  Hilbert space $\H$ of physical states.  In general, in quantum field theory, the algebra of observables in any local
region is a von Neumann algebra of Type III \cite{Araki}.   So in particular, 
 the algebra of observables in the region $P$ is a  Type III  algebra $\A$ of operators on $\H$, 
 and the algebra of observables in $P'$ is a second Type III algebra
  $\A'$, the commutant of the first ($\A'$ consists of the bounded operators that commute with $\A$, and vice-versa).
  The Type III nature of $\A$ and $\A'$  means that there is no natural notion of entropy for a state of either of these algebras.   Concretely, if one picks a state 
  $\Psi\in\H$ and attempts to compute the entropy of the state $\Psi$ reduced to a 
  region of spacetime, one will encounter an ultraviolet divergence, as first observed long ago
  \cite{Sorkin, BKLS}.
 
 \subsection{The Thermal Nature of de Sitter Space}\label{thermal} 
  
  The Hilbert space $\H$ of a quantum field theory in a fixed de Sitter background contains a distinguished state $\Psi_\dS$  \cite{CT,SS,BD,Mo,Al}, sometimes called the Bunch-Davies state.
  The Euclidean version of de Sitter space is a sphere $S^D$, and the state $\Psi_\dS$ can be defined by analytic continuation from $S^D$.
$\Psi_\dS$ is  the natural ``vacuum'' of a quantum field in a background de Sitter space; it is the analog
for de Sitter space of the Hartle-Hawking state of a black hole.
$\Psi_\dS$ is invariant under the full automorphism group $G_\dS$ of de Sitter space, which is $\SO(1,D)$ or a double cover of this to include spin. 

Once we choose to focus on a particular static patch $P$, what is relevant is not the full de Sitter automorphism group, but the subgroup $G_P$ that
consists of automorphisms of $P$.   Concretely, this subgroup is $G_P\cong \R\times \SO(D-1)$, where the first factor, which we will denote as $\R_t$, 
generates what we will call the time translations of the
static patch.    If the trajectory $\gamma$ of the observer is chosen to be the geodesic from $p$ to $q$, then $\R_t$ is the group of translations along $\gamma$, and the
second factor in $G_P$, namely $\SO(D-1)$, is the group of  rotations around $\gamma$.  
  $\R_t$ is generated by a Killing vector field $V$ that we can choose to be future-directed timelike in the
static patch $P$ and past-directed timelike in the complementary patch $P'$.   In ordinary quantum field theory in a fixed de Sitter background, $\R_t$ is generated by
a conserved charge $H$ that acts on the Hilbert space $\H$.

  As argued by Gibbons and Hawking \cite{GH} (for earlier work see \cite{FHN}), in quantum field theory in a fixed de Sitter background,
the state $\Psi_\dS$ has a thermal interpretation.   The thermal interpretation arises because, after analytic continuation
to Euclidean signature, $V$ becomes the generator of a rotation of  $S^D$.    As a result, correlation functions in the state
$\Psi_\dS$ can be analytically continued to periodic functions in imaginary time and can be interpreted 
as correlation functions in a thermal ensemble with a Hamiltonian $H_P$ and inverse temperature $\beta_\dS=2\pi r_\dS$.  Here $H_P$ generates time 
translations of the static patch and 
$r_\dS$ is the radius of curvature of the de Sitter space.  

 Importantly, $H_P$ is not the conserved charge $H$ associated to the Killing vector field
$V$ of de Sitter space; the operator $H$ has no positivity, since $V$ is past-directed in the complementary patch $P'$.   Rather, $H_P$ is supposed
to be a ``one-sided'' Hamiltonian that generates time translations in $P$, and does nothing in $P'$.   Because of fluctuations near the cosmological horizon
-- that is, near the boundary of $P$ -- such an operator actually cannot be defined in the natural
Hilbert space $\H$ of de Sitter space.   However, 
it is possible with a sort of ``brick wall'' boundary condition on the cosmological horizon to define a Hilbert space
$\H_P$ that describes excitations in region $P$ only and on which $H_P$ can be defined and is bounded below.   Correlation functions in the state $\Psi_\dS\in\H$ 
 of operators in the
patch $P$ can be interpreted as thermal correlation functions of the same operators for a thermal density matrix on $\H_P$ that is proportional to
$\exp(-\beta_\dS H_P)$.   This is one way to make precise the statement that correlation functions in the state $\Psi_\dS$ have a thermal
interpretation.  

All of this is in close analogy with the thermal nature of the Hartle-Hawking state of a 
black hole \cite{HH,I} and of the Minkowski space vacuum  as seen by an accelerated observer   \cite{Unruh}.
 A more or less equivalent  statement in the language of operator algebras is that, in the case of quantum field theory in a fixed de Sitter background,
 the ``modular Hamiltonian'' $H_\mod$ which
 generates the modular
automorphism group of the algebra $\A$ of the static patch for the state $\Psi_\dS$ is
\be\label{modham}H_\mod=\beta_\dS H,\ee
where $H$ generates time translations of the static patch.   
The analogous statement for the Rindler wedge, with the Minkowski space vacuum playing the
role of $\Psi_\dS$, is due to Bisognano and Wichman \cite{BW}; this result was carried over to black holes by Sewell \cite{Sewell},
and the de Sitter case is similar.   The interpretation of $\beta_\dS H$ as the generator of the modular automorphism group will be very important in what we say later.

What has been described so far applies to quantum field theory in a fixed de Sitter background, without dynamical gravity.   However, a number of authors
\cite{BanksOne,BanksTwo,SusskindA,Susskind}
have claimed in various ways
that this picture is substantially modified when gravity becomes dynamical, even for very small values of Newton's constant $G_N$.    Without repeating all of the arguments
here, we can motivate some of the claims as follows.   In gravity, because diffeomorphisms are gauged, conserved charges associated to diffeomorphisms of
spacetime can be computed as surface terms.   In the case of a black hole in asymptotically flat or asymptotically AdS spacetime, the energy is 
an important conserved charge, and it can be measured as a surface term at spatial infinity, namely the ADM energy.   However, the static patch in de Sitter space has
no boundary at infinity.   Its only boundary is the cosmological horizon.   Therefore, gravity will force the time generator $H_P$ of the static patch to have an interpretation
as a boundary term on the cosmological horizon.   The leading boundary term when the de Sitter radius is large is the area of the cosmological horizon.  But
according to Gibbons and Hawking, this area is an approximation to the entropy of the static patch.   So ``energy'' in the static patch is really entropy.
One argument in this direction \cite{Susskind} involves comparing the energy of a fluctuation in the static patch to the entropy
reduction associated to observing that fluctuation.   A quite different line of argument \cite{DST} that leads to a somewhat similar conclusion
involves use of a replica trick to argue that the state $\Psi_\dS$ has a flat
entanglement spectrum (so that there is no relevant notion of ``energy'' independent of entropy).   
  We will recover these claims by a simple but slightly abstract analysis of the operator
algebra of the theory.

\subsection{The Algebra of Observables}  
  
  What algebra of observables is accessible to an observer in the static patch?
A first thought might be that in ordinary quantum field theory, the observer can measure the quantum fields only in the immediate vicinity of the observer's 
worldline $\gamma$.    However, according to the Timelike Tube Theorem  \cite{Borchers,ArakiTwo}, the algebra of observables in ordinary quantum field theory in an
arbitrarily small
neighborhood of $\gamma$ is the same as the algebra of observables in the static patch.\footnote{The Timelike Tube Theorem was formulated for quantum fields
in Minkowski space.   The original proof by Borchers \cite{Borchers} used the Minkowski space structure in an essential way, but Araki's proof \cite{ArakiTwo} is based on
more robust considerations, and we expect that it carries over to a general spacetime.  In fact, the result has been proved for free field theories in spacetimes that are globally hyperbolic and real analytic \cite{Strohmaier} (the condition assumed was actually weaker than real analyticity).}   Thus, in ordinary quantum field theory, it is reasonable to claim that the observer
can access the whole algebra $\A$ of observables in $P$.

Now let us suppose that gravity is one of the fields that we want to consider in de Sitter space.   We assume, however, that $G_N$ is extremely
small, or to be more precise that the Planck length $\ell$ is much less than $r_\dS$.   Then gravity is very weakly
coupled and can be treated perturbatively.    In leading order, we make a quadratic approximation to the gravitational action and quantize gravitational perturbations
in de Sitter space in a free field approximation.   This leads to the construction of a Hilbert space $\H_\grav$ that describes gravitational fluctuations.   This must
be included as a tensor factor in defining the Hilbert space $\H$ that was the input in the previous discussion.  Thus the full Hilbert space, at this level, is
\be\label{xombo} \H=\H_{\mathrm{matt}}\otimes \H_{\mathrm{grav}} ,\ee
where now $\H_\matt$ is the Hilbert space obtained by quantizing the matter fields.

Including the weakly coupled gravitational field as one more quantum field in the construction of the Hilbert space does not qualitatively change anything what
we have said so far about the algebra of observables.    
The Type III algebra $\A$ of the static patch now  includes operators that act on the gravitational fluctuations just as they act on
any other fluctuations.     The extended Hilbert space still contains a natural state $\Psi_\dS$ that can be defined by analytic continuation from Euclidean signature
and it still has a thermal interpretation.

What does qualitatively change the picture is that, as de Sitter space is a closed universe, with compact  spatial sections, the automorphisms of de Sitter
space have to be treated as gauge constraints.  This means, in particular, that the Hilbert space that describes quantum fields and gravity in de Sitter space, in the limit
$G_N\to 0$, is not $\H$ but rather is a Hilbert space $\h\H$ that is constructed from $\H$ by imposing the de Sitter generators as constraints.  The procedure to do so
is subtle.
Naively, one might think that $\h\H$ would be the $G_\dS$-invariant subspace of $\H$, but this subspace is much too small, because, with the exception of $\Psi_\dS$,
$G_\dS$-invariant states are not normalizable.  
Instead, $\h\H$ should be defined as a space of coinvariants \cite{higuchi,marolf}.    Equivalently, as we explain in Appendix \ref{coinvariants}, one
can introduce a BRST complex for the action of $G_\dS$ on $\H$
 and define $\h\H$ as the top degree BRST cohomology. The invariant subspace is the bottom
degree BRST cohomology.

Our goal, however, is to describe not the physics of the whole de Sitter space $X$, but rather the physics of a particular static patch $P$.   The global Hilbert space
$\h\H$ is not accessible to an observer in $P$.  What is accessible to that observer is only an appropriate algebra of observables.   So what we really want to do
is to impose the constraints not on the Hilbert space but on the algebra of observables that is accessible to the observer.   Since this question depends on
a specific choice of $P$, the group of constraints that we have to impose is not the full $G_\dS$, but its subgroup $G_P$, the group of symmetries of $P$.
The important symmetry is the time translation symmetry, so we will focus on that one. 

\subsection{Including an Observer}\label{constob}

Imposing  constraints on the algebra of observables is more  straightforward conceptually than imposing constraints on the states.   As noted earlier,
to impose $G_\dS$ as a group of constraints on physical states, one does not simply require 
that a physical state should be annihilated by the group generators; the correct procedure is more subtle and is most naturally described in a BRST procedure with ghosts.  
There is no such subtlety for operators.
The subtlety in the case of physical states is possible because 
 it makes sense for the space of physical states to consist of a BRST cohomology group of states with nonzero ghost number, as long as only
one value of the ghost number is involved.
But it does not make sense 
for physical operators to carry nonzero ghost number, so in imposing a group of constraints on the algebra of operators, we can ignore the ghosts and simply
restrict to the subalgebra of invariant operators.  See Appendix \ref{coinvariants} for a fuller explanation.

In particular, in the case of the static patch, imposing  time translations as a constraint means replacing  $\A$ by $\A^H$, its subalgebra 
consisting of operators that commute with   $H$.    However, the only $H$-invariant elements of $\A$ are $c$-numbers.   
In the language of operator algebras, this is true because ``the modular automorphism group acts ergodically'' in this situation, with no nontrivial invariant operators.
Concretely, one 
might think that one could construct an $H$-invariant operator by starting with any operator $\O$ and integrating over time translations, that is, by replacing $\O$ with
\be\label{nbd}\h\O=\int_{-\infty}^\infty \d t \,e^{\i H t}\O e^{-\i H t}. \ee
However, the matrix elements of such an $\h\O$ between any two vectors $\Psi,\chi\in\h \H$ are infinite (or zero) since
\be\label{wnd} \la \Psi|e^{\i H t}\O e^{-\i H t}|\chi\ra\ee
is independent of $t$.    It will be clear from the description of $\h\H$ in Appendix \ref{coinvariants} that this is true, even though $\Psi$ and $\chi$ cannot
simply be characterized as $G_\dS$-invariant  elements of $\H$.

Since $\A^H$ is trivial, the only way to get anything sensible is to include the degrees of freedom of the observer as part of the analysis.   A minimal model of
the observer that suffices for our purposes is to say that the Hamiltonian of the observer is $H_\obs=q$, where $q$ is a new variable.   It is physically sensible to 
assume that the energy of the observer is non-negative, so we will assume that $q\geq 0$.   Thus the Hilbert space of the observer, in this model, is
$\H_\obs=L^2(\R_+)$, where $\R_+$ is the half-line $q\geq 0$.   We could also endow the observer with additional degrees of freedom,  but this would not change
anything essential in what follows.

We assume that the observer has access to any operator acting on $\H_\obs$.    Therefore, after including the observer, but prior to imposing the constraint, 
the algebra of observables is $\A\otimes B(L^2(\R_+))$, where $B(L^2(\R_+))$ is the algebra of all (bounded) operators acting on $L^2(\R_+)$.    

Now we have to impose the constraint.   The simplest model is to assume that the appropriate constraint operator is simply the sum of the Hamiltonian
$H$ of de Sitter space and the Hamiltonian of the observer:
\be\label{hsum}\h H =H+H_\obs=H+q. \ee
This is a reasonable model in the limit $G_N\to 0$, though for $G_N>0$, one would expect corrections involving positive powers of $G_N$.   
In this model, the algebra of observables, after imposing the constraint, is the $\h H$-invariant part of $\A\otimes B(L^2(\R_+))$:
\be\label{tsum}\h \A =(\A\otimes B(L^2(\R_+))^{\h H}. \ee

It turns out that this is an interesting algebra.   To analyze it, we will first ignore the condition
$q\geq 0$ and study the case that $q$ is real-valued, that is we consider the $\h H$-invariant part of the algebra $\A \otimes B(L^2(\R)$.  Let $p=-\i \d/\d q$.   We can construct by hand some operators in $\A\otimes B(L^2(\R))$ that commute with $\h H$.   One such operator is $q$ itself.
Moreover, for any $\a\in\A$, the operator $e^{\i p H}\a e^{-\i p H}$ also commutes with $\h H$.   We do not include $\h H$ itself because, as it is a constraint
operator, it annihilates physical states.    There are no other obvious operators in $\A\otimes B(L^2(\R))$
that commute with $\h H$, and a special case of Takesaki duality \cite{Takesaki} asserts that there are none.\footnote{Takesaki duality asserts an isomorphism between $\A\otimes B(L^2(\R))$ and a certain ``double crossed product'' algebra $\A_{\mathrm{dc}}$. As in the text, first define the crossed product algebra $\A_\cr$ generated by operators $e^{\i p H}\a e^{-\i p H}$
and $q$.   This algebra has an outer automorphism generated by $p$, so we can define 
 a double crossed product algebra $\A_\mathrm{dc}$ as the crossed product of the algebra $\A_\cr$ by the action of $p$. $\A_{\mathrm{dc}}$  is generated by $\A_\cr$ together with $x' +p$ where $[x',\A_\cr] = 0$. The action of $H + q$  on $\A\otimes B(L^2(\R))$ is identified under Takesaki duality with the action of translations of $x'$ on $\A_\mathrm{dc}$. Since the invariant subalgebra of $\mathcal{A}_\mathrm{dc}$ under the latter action is simply the original crossed product algebra $\A_\cr$, we conclude that $\A_\cr \cong (\A\otimes B(L^2(\R)))^{\h H}$.  See Appendix \ref{takesaki} for more background.}

Thus the invariant algebra $\A^{\h H}$ can be characterized as $\{e^{\i p H}\a e^{-\i p H},q\}$, that is, the von Neumann algebra generated by operators
$e^{\i p H}\a e^{-\i p H}$, $\a\in\A$, along with (bounded functions of) $q$.   This is actually a standard description of the ``crossed product'' of $\A$ by the 
one-parameter automorphism group generated by $H$ (see the description of $\A_L$ in section 3.1 of  \cite{GCP}), and in particular it is an algebra of Type II$_\infty$. We will denote this crossed product algebra  as $\A_\cr$.   Note that $\A_\cr$ is defined without the constraint on the observer energy.
Conjugating by $e^{-\i p H}$ leads to an equivalent description in which $\A_\cr$ is generated by operators $\a$ and $q-H$.   
In this description, the inequality $q\geq 0$ for positivity of the energy becomes $q-H\geq 0$.
In order to compare to formulas in \cite{GCP}, it is useful to define $x=-q$.   Then $\A_\cr$ is the algebra $\{\a,H+x\}$ generated by
operators $\a\in\A$ along with $H+x$; it does not matter if we take $H+x$ or $-(H+x)$ as a generator of the algebra.   In this language, the constraint
that the observer has nonnegative energy (which we have not yet imposed) is
\be\label{zolbo} H+x\leq 0. \ee

The trace in the Type II$_\infty$ algebra $\A_\cr$ can be described as follows (see \cite{GCP} for more detail).    In general, an element  $\h\a\in \A_\cr$ is an $\A$-valued function of $H+x$.
But since $H|\Psi_\dS\ra=0$, when we evaluate a matrix element $\la\Psi_\dS|\h\a|\Psi_\dS\ra$, we can set $H=0$ and view $\h\a$ as an $\A$-valued function of $x$,
which we will denote as $\a(x)$.    The trace is then\footnote{This is essentially eqn. (3.39) in \cite{GCP}, but the variable called $X$ in that equation is $\beta_\dS x$
in eqn. (\ref{tradef}).     In \cite{GCP}, the crossed product algebra was defined by adjoining $H_\mod+X$ to the bare algebra $\A$, where $H_\mod$ is the modular
Hamiltonian.   This is equivalent to adjoining $(H_\mod+X)/\beta_\dS=H+X/\beta_\dS$.   We have defined the crossed product algebra by adjoining $H+x$, so the relation
is $X=\beta_\dS x$.}
\be\label{tradef}\Tr\,\h \a =\int_{-\infty}^\infty \beta_\dS \d x\,e^{\beta_\dS x} \,\la\Psi_\dS|\a(x)|\Psi_\dS\ra.   \ee
This is well-defined and satisfies $\Tr\,\h\a\h\b=\Tr\,\h\b\h\a$ for a certain class of elements $\h\a,\h\b\in\A_\cr$.   
It is also positive in the sense that $\Tr\,\h\a^\dagger\h\a>0$ for all $\h\a\not=0$.   
 But it is not well-defined for all elements;
for example, the trace of the identity element of $\A_\cr$ is divergent.   The reason that this happened is that we have implicitly assumed $\Psi_\dS$ to be 
independent of $x$.   However, a constant function  on $\R$ is not  square-integrable, so $\Psi_\dS$ extended in this way is not an element of $\H\otimes L^2(\R)$.
Thus the trace defined in eqn. (\ref{tradef}) is not a ``state'' of $\A_\cr$  but a  ``weight.''   Here, a  state on an algebra 
is a linear function $\h\a\to F(\h\a)$  that is positive in the sense that
$F(\h\a^\dagger \h\a)\geq 0$ for all $\h\a\not=0$; a weight is precisely the same, except that it is not defined for all elements of the algebra (it equals
$+\infty$ for some elements).

Now we want to impose the constraint that $q\geq 0$.   To do this, let $\Theta(q)$ be the function that is 1 for $q\geq 0$ and 0 for $q<0$.   Multiplication by
$\Theta(q)$ is a projection operator $\Pi$, acting on $\H\otimes L^2(\R)$.   We can incorporate the constraint by just replacing the algebra $\A_\cr$ with
\be\label{repalg}\h\A=\Pi \A_\cr \Pi. \ee
In other words, the operators are the same as before, but restricted to act between states that are in the image of $\Pi$.   $\h\A$ can be viewed as a von Neumann
algebra acting on the Hilbert space $\Pi(\H\otimes L^2(\R))=\H\otimes L^2(\R_+)$.   $\h\A$ automatically comes with a trace, namely the restriction of the trace
on $\A_\cr$ to operators of the form $\Pi\h\a\Pi$.  

An infinite-dimensional von Neumann algebra that has a trace that takes a finite value for the identity element is of Type II$_1$.   So to show that $\h\A$ is of
Type II$_1$, it suffices to show that the trace in this algebra takes a finite value for the identity element.
The identity element of $\h\A$ corresponds to the element $\Pi=\Theta(q)=\Theta(-H-x)$ of $\A_\cr$, so we compute
\be\label{comdel}\Tr_{\h\A}\,1 =\Tr_{\A_\cr}\,\Pi=\int_{-\infty}^\infty \beta_\dS\d x\,e^{\beta_\dS x} \la\Psi_\dS|\Theta(-H-x) |\Psi_\dS\ra =\int_{-\infty}^0 \beta_\dS \d x \,e^{\beta_\dS x}=1. \ee
Thus the algebra of observables in de Sitter space including the observer is of Type II$_1$,  and moreover
the trace that we have defined is normalized so that $\Tr\,1=1$.

 It is also true on very general grounds  that $\h\A=\Pi\A_\cr\Pi$ is a factor, meaning that its center consists only of the complex scalars.
   In general, if $\B$ is a von Neumann algebra that is a factor
and $\Pi$ is a projection operator in $\B$, then the von Neumann algebra $\Pi\B\Pi$ is also a factor.\footnote{The algebra $\Pi\B\Pi$  acts on $\Pi\H$, and its center is the intersection 
of $\Pi\B\Pi$ with its commutant.   It is shown in \cite{VJNotes}, in statement EP7) proved on p. 21, that the commutant of $\Pi\B\Pi$ is $\Pi\B'$, where $\B'$
is the commutant of $\B$ in $\H$.   So the center of $\Pi\B\Pi$ is the intersection of $\Pi\B\Pi$ with $\Pi\B'$.    If $\B$ is a factor, which means that the intersection of
$\B$ and $\B'$ consists only of $\C$, then the intersection of $\Pi\B\Pi$ and $\Pi\B'$ consists only of multiples of the identity of $\Pi\B\Pi$, 
 and therefore $\Pi\B\Pi$ is a factor.}

A Type II$_1$ algebra that is ``hyperfinite,'' meaning that it can be approximated by finite-dimensional matrix algebras, is isomorphic to the Murray-von Neumann
algebra that was described in section \ref{whatis}.    Algebras of local regions in quantum field theory -- and their crossed products with finite-dimensional automorphism
groups --- are believed to be always hyperfinite.  So we expect that $\h\A$ is isomorphic to the algebra, described in section \ref{whatis}, that acts on
an infinite collection of qubits in an almost maximally mixed state.

As we explained in section \ref{whatis}, the trace in a Type II$_1$ algebra is defined for all elements of the algebra, and it is possible for a state of such an algebra
to introduce density matrices and entropies that share many properties with density matrices and entropies in ordinary quantum mechanics.
We also explained that a Type II$_1$ algebra has a state of maximum entropy, namely the state with density matrix $\rho=1$.

To understand what is the state of maximum entropy in the case of the algebra $\h\A$, we can compute expectation values in this state.   First consider an operator $\a\in \A$.
  For $\rho=1$, the expectation value of $\a$ is
  \be\label{donbo}\Tr\,\a\rho =\Tr\,\a=\int_{-\infty}^0 \beta_\dS \d x \,e^{\beta_\dS x} \la\Psi_\dS|\a|\Psi_\dS\ra=\la\Psi_\dS|\a|\Psi_\dS\ra. \ee
  On the other hand, consider an operator of the form $G(-H-x)$,  where $G$ is some bounded function.   Bearing in mind that $H|\Psi_\dS\ra=0$, $\la\Psi_\dS|  \Psi_\dS\ra=1$, and $x=-q$, we get   \be\label{wonbo}\Tr\, G(-H-x)\rho=\Tr\, G(-H-x)=\int_{-\infty}^0\beta_\dS \d x e^{\beta_\dS x}\la\Psi_\dS|G(-H-x)|\Psi_\dS\ra=\int_0^{\infty}\beta_\dS \d q \,e^{-\beta_\dS q} G(q). \ee
  Thus we can think of the maximum entropy state as the ordinary de Sitter state $\Psi_\dS$ of the quantum fields, tensored with a thermal energy distribution $p(q) = \beta_\dS \,e^{-\beta_\dS q}$ for the observer.
  More formally, the maximum entropy state has the following purification: the Hilbert space is $\H\otimes L^2(\R_-)$ (where $\R_-$ is the half-line $x\leq 0$ and
  $\h\A$ acts as described earlier),
  and the state is 
  \be\label{pureds} \Psi_\max=\Psi_\dS \, \sqrt{\beta_\dS} \,e^{\beta_\dS x/2}.\ee   We will discuss this purification further in section \ref{hilbertspaces}.  
  Since the maximally entropic state has density matrix $\rho_\max=1$, we have for any $\x\in\h\A$
  \be\label{funfact}\Tr\,\x=\Tr\, \x\rho_\max=\la\Psi_\max|\x|\Psi_\max\ra. \ee
  
R\'{e}nyi entropies can be defined as usual as $S_\alpha(\rho)=\frac{1}{1-\alpha}\log \Tr\,\rho^\alpha.$
Clearly, all R\'{e}nyi entropies vanish for the maximally entropic state with $\rho=1$; thus, this state is analogous to a state in ordinary quantum mechanics that
has a flat entanglement spectrum, and therefore has R\'{e}nyi entropies that are independent of $\alpha$.
   That the state $\Psi_\dS$ has a flat entanglement spectrum was argued by Dong, Silverstein, and Torroba via a Euclidean path
integral and a replica trick \cite{DST}.  Their reasoning can be extended to include the observer, as we discuss in section \ref{hilbertspaces}.

The flat entanglement spectrum also implies that  after coupling to gravity, the suppression of fluctuations in de Sitter space can be understood purely in entropic terms, 
 as advocated in \cite{BanksOne,BanksTwo,Susskind}. 
Let $\Lambda\in\h\A$ be a projection operator, and suppose that the observer performs an experiment in which the possible outcomes correspond to $\Lambda$
and $1-\Lambda$.   In the maximally entropic state with density matrix $\rho=1$, the probability of the 
outcome corresponding to $\Lambda$ is $p_\Lambda=\Tr\,\Lambda$.   After that outcome is
observed, the system can be described by the density matrix $\rho_\Lambda=\Lambda/\Tr\,\Lambda$.     The von Neumann entropy of the new density matrix
is $S(\rho_\Lambda)=-\Tr\,\rho_\Lambda\log \rho_\Lambda=-\log(1/\Tr\,\Lambda)$, as computed in section \ref{whatis}.
$S(\rho_\Lambda)$ is the same
as the entropy deficit $\Delta S_\Lambda=S(\rho_\Lambda)-S(\rho)$ associated with the observed outcome, since $S(\rho)=0$.
Thus we have arrived at a rather abstract explanation of the relation
\be\label{zolgo} p_\Lambda = e^{\Delta S_\Lambda} \ee
that was claimed in \cite{BanksOne,Susskind}.

After incorporating the observer and imposing the constraints, an operator $\a$  is replaced by its dressed version $\h \a =\Pi e^{\i p H}\a e^{-\i p H}\Pi$, where $\Pi=\Theta(q)$
is the projection operator onto states in which the observer energy is non-negative.   The expectation value of $\h\a$ in the maximum entropy state $\Psi_\max$ is the
same as the expectation value of $\a$ in the de Sitter state $\Psi_\dS$:
\be\label{compar} \la \Psi_\max|\h\a|\Psi_\max\ra=\la\Psi_\max|\Pi e^{\i p H}\a e^{-\i p H}\Pi|\Psi_\max\ra =\la\Psi_\dS|\a|\Psi_\dS\ra,\ee
since $H\Psi_\max=0$ and $\Pi\Psi_\max=\Psi_\max$.    Note that here $\a$ is a completely general element of the algebra of observables, not
necessarily a local operator.   For example, $\a$ can be a product of local operators at different times, $\a=\a_1(t_1)\a_2(t_2)\cdots \a_n(t_n)$.

\subsection{What Is An Observer?}\label{observer}

Introducing an observer was necessary to give a sensible result in the preceding analysis, but may seem artificial.  In a satisfactory
theory, we are not entitled to introduce an observer from outside.  The observer should be described by the theory.

Our requirement for what an observer should  be is quite minimal. The role of the observer was to help us fix the time translation symmetry of de Sitter
space, so an observer is any system that can tell time.   We chose a simple model
in which a complete set of  commuting observables in the observer's Hilbert space is the Hamiltonian $H_\obs=q$.   This is not necessary; we could
endow the observer with additional operators that commute with $H_\obs$.  In fact,  our model was unrealistically simple;  in a more realistic
model, we would at least want to describe the position of the observer in de Sitter space.   However, endowing the observer with operators that commute with
$H_\obs$ would not affect the analysis in an interesting way.   Such operators would just go along for the ride.  

One can explain as follows the role of the observer in our analysis.   Let $\h\H$ be the Hilbert space of de Sitter space after imposing the gravitational
constraints, as reviewed in Appendix \ref{coinvariants}.   This Hilbert space exists, and there are operators that act on it.  But  no operators on
$\h\H$  can be defined just in the static patch.   Therefore, rather than all of $\h\H$, one considers a ``code subspace'' consisting of states in $\h\H$
in which the static patch contains an observer with some assumed properties.    There are operators in the static patch that are well-defined on the code subspace,
though these operators are not well-defined on all of $\H$.   What we have studied is the algebra of operators in the static patch that act on such a code subspace.

For a rather non-minimal model of an observer in de Sitter space, we could take the Local Group of galaxies that are gravitationally bound to the Milky Way.
Assuming that the accelerating expansion of the present universe is the beginning of a phase of exponential expansion in de Sitter space, within roughly
$10^{11}$ years galaxies that are not gravitationally bound to the Milky Way will be behind a cosmological horizon.   The Local Group will persist, with only
relatively slow changes, for a time vastly longer than the time scale of the de Sitter expansion.   An analysis similar to what we have presented (but taking into
account the many degrees of freedom of the Local Group) is applicable to a code subspace of states in which the Local Group is present in de Sitter space.

A minimal modification of the model would be to give the observer a mass $m\gg T_\dS$, and replace $q\geq 0$ with $q\geq m$.  This, together with appropriate
boundary conditions, would give a rationale for assuming that the observer worldline  is localized along the geodesic at the center of the static patch.   The algebra of
observables is still of Type II$_1$.

\subsection{Gravitational Dressing, and Giving the Observer an Orthonormal Frame}\label{frame}

The algebra $\h\A$ that we have defined for the static patch with an observer present might be described as an algebra of gravitationally dressed operators.   In the
static patch of de Sitter space, there is no region at spatial infinity to which an operator could be gravitationally dressed, so instead $\h\A$ is an algebra of operators
that have been gravitationally dressed to the observer.   

We only discussed explicitly the group $\R_t$ of time translations of the static patch, and we have not taken into account the second factor
in the static patch symmetry group $G_P=\R_t\times \SO(D-1)$.   This second factor is the rotation group of the static patch.   Since the group $\SO(D-1)$ is
compact, imposing  $\SO(D-1)$ as a group of constraints simply means requiring that operators should be $\SO(D-1)$-invariant.   Though there is nothing wrong
with this, one loses a great deal of information if one is only able to gravitationally dress the rotation-invariant operators.   As an alternative, we could equip
the observer with an orthonormal frame, as well as a Hamiltonian.   At the classical level, this means that the phase space of the observer would be not
$T^*\R_+$, where $\R_+$ is the half-line $q\leq 0$, but $T^*\R^+\times T^*\SO(D-1)$.   With such a model of the observer, we would be able to gravitationally
dress all operators in the static patch, not just the ones of zero angular momentum.   The resulting algebra is still of Type II$_1$.  

In case the observer is the Local Group of galaxies, as discussed in section \ref{observer}, this step is unnecessary as the Local Group is not invariant under
any nontrivial rotations, so arbitrary operators in the static patch could be gravitationally dressed to the Local Group.

Equipping the observer with an orthonormal frame, or some other mechanism that breaks the rotation symmetry, is actually important in section \ref{bulk},
because we will  assume that all operators in $\A$, not just the ones that commute with rotations, can be gravitationally dressed to the observer.   Otherwise
we would not get the standard relative entropy.    However, for simplicity, the analysis in section \ref{bulk} is written without introducing an explicit symmetry-breaking
mechanism.

\section{A Bulk Formula For The Entropy}\label{bulk}

We have argued that the operators accessible to an observer in de Sitter space form a von Neumann algebra $\h \A$ of Type II$_1$.  To a state of such an algebra,
one can associate an entropy.   The goal of the present section is to describe a bulk formula for the entropy of a state of $\h\A$ that is semiclassical in a sense
that we will describe.   (This discussion is in close parallel with a treatment of the black hole in a companion paper \cite{CPW}.) 

As in section \ref{constob}, $\h\A=\Pi \A_\cr \Pi$, where $\A_\cr$ is the crossed product algebra generated by $\A$ and $H+x$ and $\Pi$ is the projection operator
$\Pi=\Theta(-H-x)$.   $\A_\cr$ can act on the Hilbert space $\h\H=\Pi(\H\otimes L^2(\R))$, where $\A$, $H$ act on $\H$ and $x$ acts on $L^2(\R)$.   We will
explain in section \ref{hilbertspaces} that every state of the algebra $\h\A$ -- that is, every density matrix $\rho\in\h\A$ -- can be purified by a pure state in $\h\H$.
(We will also explain a natural setup in which $\h\H$ emerges as the space of physical states.)   For now, we can just think of a choice of a state $\h\Phi\in\h\H$
as a convenient way to describe a state of the algebra $\h\A$.

We will consider states of the form 
\be\label{newstate}\h\Phi =\Phi\otimes f(x) ,\ee
with $\Phi \in\H$, $f(x)\in L^2(\R)$.  
Because of the projection operator $\Pi$, it is natural to assume that the function $f(x)$ has support for $x<0$.   We assume a normalization condition
\be\label{normcon}\la\Phi|\Phi\ra =1 =\int_{-\infty}^0\d x \,|f(x)|^2. \ee
We want to impose a further condition on $f$ that will ensure 
roughly that the full spacetime, including the observer, is a definite semiclassical spacetime in which, when a given event is occurring, the observer's clock
shows a well-defined time, with the uncertainty in time being much less than $\beta_\dS$.
  We recall that, before conjugation by $e^{-\i p H}$, the Hamiltonian of the observer is $H_\obs =q=-x$.  Prior to imposing the constraint
$q\geq 0$, there is a self-adjoint operator $p$ conjugate to $q$ with $[p,q]=-\i$, so $\i [H_\obs,p]=-1$.   This tells us that $-p$ is the time told by the observer's 
clock; since $[e^{-\i p H},p] = 0$, this continues to be true in the conjugated description.\footnote{Once one takes into account the constraint $q\geq 0$, a self-adjoint operator $p$ obeying $[p,q]=-\i$ does not exist, so quantum mechanically, a clock
whose energy is bounded below cannot tell time perfectly.  But it can tell time very well, for a long period.  For example, $p$ can be defined for states whose
support is away from $q=0$, so $p$ is approximately well-defined for any state that is supported mostly away from $q=0$, 
such as the states considered in the text.}  We would like the observer to be able to measure the
times at which events occur in de Sitter space with a precision much greater than the natural de Sitter time scale $\beta_\dS$.   For this, the function $f(x)$
should be slowly varying.   We choose
\be\label{zimbo} f(x) =\epsilon^{1/2} g(\epsilon x), ~~~\epsilon\ll \beta_\dS. \ee
where $g(x)$ is a smooth, bounded function with support for $x<0$ and  $\epsilon$ is a small parameter.   We assume a normalization condition
$\int_{-\infty}^0\d x \,|f(x)|^2=1$.
In a state of this kind, 
$p\approx 0$, with an uncertainty of order $\epsilon$.  After this state evolves for a time $t$ with Hamiltonian $H_\obs=q$,
 it has $p\approx -t$, with the same uncertainty.
Such a  function  $f(x)$  is mostly supported for $x\sim -{1}/{\epsilon} \ll0$, and hence 
 $\Phi\otimes f(x)$ is approximately invariant under the projection operator
 $\Pi=\Theta(-H-x)$.  So we can view $\h\Phi=\Phi\otimes f(x)$ as an element of $\h\H=\Pi (\H\otimes L^2(\R))$.   
 
 An important and slightly perplexing detail is that the maximum entropy state $\Psi_\max=\Psi_\dS \otimes \sqrt{\beta_\dS} e^{\beta_\dS x/2}$ is not
 a semi-classical state in this sense; in this state, $p$ has an uncertainty of order $\beta_\dS$.   The following analysis of the density matrix of a state of the form
 $\h\Phi$ therefore does not apply to $\Psi_\max$.  (The density matrix of $\Psi_\max$ is the identity operator, as explained in section \ref{constob}.)

  To compute the entropy of the state $\h\Phi$, we will first find an approximate formula for its density matrix $\rho_{\h\Phi}$ and then evaluate the von Neumann
  entropy $S(\rho_{\h\Phi})=-\Tr\,\rho_{\h\Phi}\log\,\rho_{\h\Phi}$.   
 Let $\Psi_\dS\in\H$ be the natural Bunch-Davies state, and let $\Delta_{\Psi_\dS}:\H\to\H$
 be its modular operator for the algebra $\A$.   To slightly shorten the formulas
in the following derivation, we will write just $\Psi$ and $\beta$ for $\Psi_\dS$ and $\beta_\dS$.

    We have $\Delta_{\Psi}=e^{-h_{\Psi}}$ where $h_{\Psi}$ (which was called $H_\mod $ in section 
 (\ref{thermal})) is often called the modular Hamiltonian.
 We will also need the corresponding relative modular operator 
 $\Delta_{\Phi|\Psi}:\H\to\H$ for the algebra $\A$ and the states $\Phi$ and $\Psi$.   
 This operator is defined by $\Delta_{\Phi|\Psi}=S^\dagger_{\Phi|\Psi} S_{\Phi|\Psi}$,
 where $S_{\Phi|\Psi}$ is the relative Tomita operator, which is antilinear and  satisfies $S_{\Phi|\Psi}\a\Psi=\a^\dagger\Phi$, for $\a\in\A$.
 From this it follows that\footnote{In the step $ \la\Psi|S^\dagger_{\Phi|\Psi}\a^\dagger|\Phi\ra= 
  \overline{\la\Phi|\a^\dagger|\Phi\ra}$, we use the fact that $S_{\Phi|\Psi}$ is antilinear and $S_{\Phi|\Psi}|\Psi\ra=|\Phi\ra.$}
  \be\label{welgo}\la\Psi|\Delta_{\Phi|\Psi}\a|\Psi\ra =\la\Psi|S_{\Phi|\Psi}^\dagger S_{\Phi|\Psi}\a|\Psi \ra= \la\Psi|S^\dagger_{\Phi|\Psi}\a^\dagger|\Phi\ra= 
  \overline{\la\Phi|\a^\dagger|\Phi\ra}=\la\Phi| \a|\Phi\ra,~~~\a\in\A. \ee    
We write $\Delta_{\Phi|\Psi}=e^{-h_{\Phi|\Psi}}$, where $h_{\Phi|\Psi}$ is the relative modular Hamiltonian.   

The desired density matrix $\rho_{\h\Phi}$ is supposed to satisfy \be\label{supposed} \la\h\Phi|\h\a|\h\Phi\ra=\Tr\,\rho_{\h\Phi}\h\a, ~\h\a\in\h\A.\ee
   But $\Tr\,\rho_{\h\Phi}\h\a
=\la\Psi_\max|\rho_{\h\Phi}\h\a|\Psi_\max\ra$, according to eqn. (\ref{funfact}).   So the condition we want is
\be\label{goodcond} \la\Psi_\max|\rho_{\h\Phi} \h\a|\Psi_\max\ra =\la\h\Phi|\h\a|\h\Phi\ra. \ee
There is an obvious similarity between  eqns. (\ref{welgo}) and (\ref{goodcond}), suggesting that $\rho_{\h\Phi}$ can be constructed in a simple way by modifying
$\Delta_{\Phi|\Psi}$ to include $x$.   In doing so, we have to remember that as well as satisfying (\ref{goodcond}), $\rho_{\h\Phi}$ is supposed to be an element
of $\h\A$ (or possibly an unbounded operator affiliated to $\h\A$, meaning that bounded functions of $\rho_{\h\Phi}$ are in $\h\A$).   The operator $x$ is not an element
of $\h\A$, but a general bounded function of $x+h_\Psi/\beta$ is in $\h\A$.   Also, $\Delta_{\Phi|\Psi}$ itself is not affiliated to $\h\A$, but as we will discuss
shortly, $e^{-\beta x}\Delta_{\Phi|\Psi}$ is affiliated to $\h\A$.

Taking these facts into account, one can find an approximation to the density matrix:
\be\label{densmat}\rho_{\h\Phi}=\frac{1}{\beta}\bar f(x+h_{\Psi}/\beta)e^{-\beta x}\Delta_{\Phi|\Psi}  f(x+h_{\Psi}/\beta) +\O(\epsilon). \ee
First of all, the operator $\O= \bar f(x+h_{\Psi}/\beta)e^{-\beta x}\Delta_{\Phi|\Psi}  f(x+h_{\Psi}/\beta)$ is manifestly self-adjoint and non-negative.
The operators   $f(x+h_{\Psi}/\beta)$, $\bar f(x+h_{\Psi}/\beta)$ are elements of $\h\A$, since they are bounded functions of $x+h_\Psi/\beta$.   We have
\be\label{turno} e^{-\beta x} \Delta_{\Phi|\Psi}= e^{-(\beta x+h_{\Psi})+(h_{\Psi}-h_{\Phi|\Psi})}.\ee
One term in the exponent is the algebra  generator $\beta x+h_{\Psi}$.
To understand the other term, we use the Connes cocycle, defined as
\be\label{toro} u_{\Phi|\Psi}(s)=\Delta_{\Phi|\Psi}^{\i s} \Delta_\Psi^{-\i s}= \Delta_\Phi^{\i s}\Delta_{\Psi|\Phi}^{-\i s}. \ee
The important properties of $u_{\Phi|\Psi}(s)$ for our purposes are that for real $s$ it is valued in $\A$, and that the two formulas for $u_{\Phi|\Psi}(s)$ are in fact
equal.   See for example section 6 of \cite{Lashkari}.   Differentiating the first formula for $u_{\Phi|\Psi}(s)$ with respect to $s$
at $s=0$, we learn that $h_\Psi-h_{\Phi|\Psi}$ is affiliated to $\A$.   
  So 
the operator  in eqn. (\ref{turno}) is affiliated with $\h\A$.  The equivalence between the two formulas in eqn. (\ref{toro}) implies,
again by differentiating with respect to $s$ at $s=0$, that
\be\label{longid} h_{\Phi|\Psi}-h_\Psi =h_\Phi-h_{\Psi|\Phi}. \ee

In verifying eqn. (\ref{goodcond}), 
we will make use of the fact that $f(x)$ is slowly varying, which implies that $f(x+h_{\Psi}/\beta)$ approximately commutes with
$\Delta_{\Phi|\Psi}$.    Hence instead of eqn. (\ref{densmat}), we can equivalently write
\be\label{densermat}\rho_{\h\Phi}=\frac{1}{\beta} |f(x+h_{\Psi}/\beta)|^2e^{-\beta x}\Delta_{\Phi|\Psi}  +\O(\epsilon). \ee

It suffices to verify eqn. (\ref{goodcond}) for $\h\a=\a e^{\i u(\beta x+h_\Psi)}$, $\a\in\A$, $u\in \R$, 
since any element of $\h\A$ can be approximated by a sum of such operators.   We have chosen
the state $f(x)$ of the observer's clock so that $|p|\lesssim \epsilon$.   But multiplication by $e^{\i u \beta x}$ shifts $p$ by $-u\beta$.   As a result,
the left and right hand sides of eqn. (\ref{goodcond}) vanish exponentially unless $|u|\lesssim \epsilon/\beta$, so we can restrict to that range of $u$.
We have
\be\label{oppo}\la\h\Phi|\h\a|\h\Phi\ra= \int_{-\infty}^0 \d x \,|f(x)|^2 \la\Phi|\a e^{\i u(h_\Psi+\beta x)} |\Phi\ra.\ee
For $|u|\lesssim \epsilon/\beta$, we can drop the factor $e^{\i u h_\Psi}$, with an error of order $\epsilon$.
Once this is done, the integrand in eqn. (\ref{oppo}) involves a matrix element $\la\Phi|\a|\Phi\ra$ with $\a\in \A$.
Hence we can use eqn. (\ref{welgo}), giving
\be\label{noppo}\la\h\Phi|\h\a|\h\Phi\ra= \int_{-\infty}^0 \d x \,|f(x)|^2 e^{\i u \beta x} \la\Psi |\Delta_{\Phi|\Psi} \a |\Psi\ra.\ee
Since $h_\Psi|\Psi\ra=0$, we can move any function of $x$ inside the matrix element and replace $x$ by $x+h_\Psi/\beta$:
\be\label{zoppo} \la\h\Phi|\h\a|\h\Phi\ra = \int_{-\infty}^0 \d x \,  \left\la\Psi \left|\,| f(x+h_\Psi/\beta)|^2 \Delta_{\Phi|\Psi} \h\a \right|\Psi\right\ra.\ee
Multiplying and dividing by $\beta e^{\beta x}$, we get 
\be\label{twofo}  \la\h\Phi|\h\a|\h\Phi\ra
=\int_{-\infty}^0\d x\,\beta e^{\beta x} \left\la\Psi \left|\frac{1}{\beta} |f(x+h_\Psi/\beta)|^2 e^{-\beta  x}\Delta_{\Phi|\Psi} \h\a\right |\Psi\right\ra \ee
 \be\notag~~~~=\left\la\Psi_\max\left| \frac{1}{\beta}  |f(x+h_\Psi/\beta)|^2 e^{-\beta  x}\Delta_{\Phi|\Psi}\h\a \right|\Psi_\max\right\ra.\ee
 Comparing to eqn. (\ref{densermat}),  this establishes the claimed formula for the density matrix.
 
 The entropy is $S(\rho_{\h\Phi}) =-\Tr\,\rho_{\h\Phi}\log \rho_{\h\Phi}=-\la\h\Phi|\log\rho_{\h\Phi}|\h\Phi\ra. $   Because $f(x)$ is slowly varying and supported
 mostly at large $|x|$, in evaluating $\log \,\rho_{\h\Phi}$, we can approximate $f(x+h_\Psi/\beta)$ by $f(x)$.   We get then 
 \be\label{logform}-\log \,\rho_{\h\Phi}= h_{\Phi|\Psi} +\beta x - \log |f(x)|^2 +\log \beta.\ee
 The entropy then is 
 \be\label{entans} S(\rho_{\h\Phi})=\la\Phi|h_{\Phi|\Psi}|\Phi\ra+\int_{-\infty}^0\d x |f(x)|^2 (\beta x-\log |f(x)|^2 +\log \beta).\ee
 \be\notag   ~~~~~~~~~~~~~~~~~~~=\la\Phi|h_{\Phi|\Psi}|\Phi\ra +\la \h\Phi|\beta x|\h\Phi\ra +\int_{-\infty}^0\d x |f(x)|^2(-\log |f(x)|^2+\log\beta).\ee
 It is convenient, however, to use the identity of eqn. (\ref{longid}).   Since $\la\Phi|h_\Phi|\Phi\ra=0$, we get
\be\label{betterid} S(\rho_{\h\Phi}) =-\la\Phi|h_{\Psi|\Phi}|\Phi\ra +\la \h\Phi|h_\Psi+\beta x|\h\Phi\ra +\int_{-\infty}^0\d x |f(x)|^2(-\log |f(x)|^2+\log\beta).\ee

 We now want to show that this formula, which has been derived based on rather abstract considerations, agrees with what one would expect from gravity.
 To be more precise, since the entropy of a Type II$_1$ algebra is a renormalized entropy whose definition involves a subtraction, we will only reproduce
 the expected results from gravity up to an overall additive constant, independent of the state.   
 One can think of this constant as the entropy of the maximum entropy state, which in the
 Type II$_1$ algebra is defined to be 0.
 
\begin{figure}
 \begin{center}
   \includegraphics[width=2.2in]{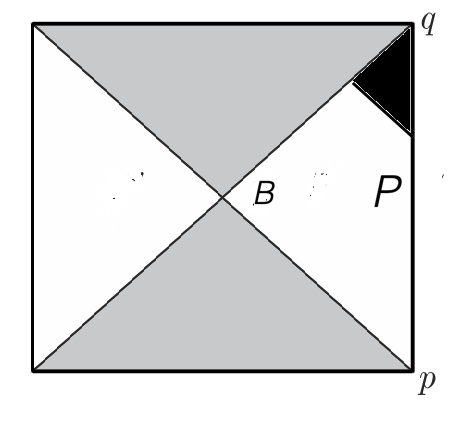}
 \end{center}
\caption{\small In this diagram, the ``exterior'' region for a given cut of the future horizon is the region to the right of, and spacelike separated
from, the given cut.
In the case of a cut along the surface $B$ -- the intersection of the past and future horizons
of the static patch $P$ -- the exterior region is simply $P$.  
What is exterior to a later cut of the future horizon is a smaller spacetime region, such as the example shaded in black. \label{DDD}}
\end{figure}

First of all, following Bekenstein \cite{Bekenstein}, the generalized entropy of a horizon is defined as 
\be\label{intox} S_\gen=\frac{A}{4 G_N}+S_\out , \ee
where $A$ is the horizon area and $S_\out $ is the entropy of the fields exterior to the horizon.  To be more exact, one can evaluate this formula for any spacelike
surface that is a codimension 1 ``cut'' of the horizon.   $S_\gen$ is then supposed to be the entropy (including gravitational entropy) of the region exterior to (and spacelike
separated from)   the given cut.   In the case of de Sitter space, if we choose the cut to be the ``bifurcate horizon'' $B$ (the intersection of past and future horizons, 
sketched in fig. \ref{DDD}), then $S_\gen$ is expected
to be the entropy of the static patch.    If we choose a later cut, then $S_\gen$ is an entropy of a smaller spacetime region, as also illustrated
in the figure.

The generalized entropy for different horizon cuts can be compared by an elegant formula \cite{Wall}.   A particularly simple and useful case of this formula
compares $S_\gen(B)$, the generalized entropy of the bifurcate horizon $B$, to $S_\gen(\infty)$, the limit of the generalized entropy as the horizon cut goes
to future infinity.   In this case, the formula reads\footnote{We elaborate upon the derivation of this formula in a companion paper \cite{CPW}.}
\be\label{formreads} S_\gen(B) =S_\gen(\infty)- S_\rel(\Phi||\Psi). \ee  
Here $S_\rel(\Phi||\Psi)$ is the relative entropy between a state $\Phi$ of the bulk fields and the natural de Sitter state $\Psi=\Psi_\dS$.   The state
$\Psi_\dS$ (or the Hartle-Hawking state, in the case of a black hole) enters
the derivation because its modular Hamiltonian generates time translations, which act as boosts of the horizon, leaving fixed the cut $B$.   

To compare these formulas, the first step is to observe that one term in eqn. (\ref{entans}) is the relative entropy in eqn. (\ref{formreads}):
\be\label{goodnews} -S_\rel(\Phi||\Psi) =\la\Phi|\log\Delta_{\Psi|\Phi}|\Phi\ra=-\la\Phi|h_{\Psi|\Phi}|\Phi\ra.\ee
This is indeed Araki's formula for relative entropy in the general context of von Neumann algebras \cite{ArakiThree}.   In the present discussion, $\Phi$ and $\Psi$
are states of the underlying Type III$_1$ algebra $\A$ of quantum fields in de Sitter space, and Araki's formula for relative entropy is essentially the only one available.
In the more familiar case of an algebra of Type I, with states $\Psi,\Phi$ corresponding respectively to density matrices $\sigma,\rho$, one
can show that $-\la\Phi|\log\Delta_{\Psi|\Phi}|\Phi\ra =\Tr\,\rho\log\rho-\Tr\,\rho\log\sigma$, where the last formula  is a possibly  more familiar
definition of relative entropy.\footnote{The equivalence of the two formulas for relative entropy in the case of
an algebra of Type I is shown, for example, in section 4.3 of \cite{WittenNotes}.
(See footnote 16 of that paper for the conventions used there for the relative modular operator.)}
 
 So to reconcile our formula (\ref{betterid}) with the expected result (\ref{formreads}), we need 
 \be\label{urgo}S_\gen(\infty)\overset{?}{=} \la \h\Phi|h_\Psi+\beta x|\h\Phi\ra +\int_{-\infty}^0\d x |f(x)|^2(-\log |f(x)|^2+\log\beta)+{\mathrm{constant}},\ee
 where we include in the formula an additive constant  which is not captured by the Type II$_1$ algebra.
 Of course, the generalized entropy in the far future is supposed to be
 \be\label{nurgo}S_\gen(\infty)=\frac{A(\infty)}{4G_N} + S_\out(\infty),\ee
 where $A(\infty)$ and $S_\out(\infty)$ are the horizon area and the entropy outside the horizon in the far future.   
 
 Because of the exponential expansion of de Sitter space, typical perturbations cross the horizon and disappear from the static patch in a time of order $\beta$.
 Hence after a few times $\beta$, the static patch contains an observer in empty de Sitter space.  It follows that $S_\out(\infty)$ is just the entropy of the observer.   
 In our model, we have attributed  to the observer no property except an energy, and therefore the entropy of the observer comes only from fluctuations in energy.
 We can interpret the term $\int_{-\infty}^0\d x |f(x)|^2(-\log |f(x)|^2+\log\beta)$ on the right hand side of eqn. (\ref{urgo}) as the entropy in the observer's energy
 fluctuations.   The time at which this entropy is measured does not matter, since in the model, the observer energy $H_\obs=q=-x$ is a conserved quantity.
 So we identify $\int_{-\infty}^0\d x |f(x)|^2(-\log |f(x)|^2+\log\beta)$ as $S_\out(\infty)$.   
 
 It remains to understand the term $\la\h\Phi|h_\Psi+\beta x|\h\Phi\ra$ in the formula for the entropy.   Recall that the conjugation by $e^{-\i p H}$ mapped $H_\obs = q = -x$ to $-x - H$. It follows that
\be
\la\h\Phi|h_\Psi+\beta x|\h\Phi\ra = - \beta \la H_\obs \ra.
\ee
 To complete the analysis, one needs the fact that the presence in the center of the static patch of an object of energy $E$, assuming that $E$ is small enough
 that one can work to linear order in $E$, reduces $A_\hor/4G_N$, where $A_\hor$ is the area of the cosmological horizon, by $\beta E$ \cite{Susskind}.
 In four dimensions, for example, the Schwarzschild-de Sitter
 metric for the case of an object of energy $E$ at the center of the static patch is $\d s^2=-f(r) \d t^2+\frac{1}{f(r)}\d r^2+r^2\d\Omega^2$, with
 \be\label{metfac} f(r) = 1-\frac{2 GE}{r} -\frac{r^2}{r_\dS^2}, \ee
 where the de Sitter radius is $r_\dS=\beta/2\pi$.   The cosmological horizon is at the zero of $f(r)$ that approaches $r_\dS$ for $E\to 0$.
 This is at
 \be\label{etfac}r_\hor =r_\dS-GE+\O(E^2). \ee
 The horizon area is $A_\hor=4\pi r_\hor^2$, so
 \be\label{neffac}\frac{A_\hor}{4 G_N}= \frac{\pi r_\dS^2}{ G_N}-2\pi E r_\dS+\O(E^2)=\frac{\pi r_\dS^2}{G_N}-\beta E    +\O(E^2).\ee
 This shows the claimed shift $-\beta\la H_\obs\ra$ in the entropy of the horizon due to an object at the center of the static patch.
 
 We can therefore interpret the term $-\beta\la\h\Phi|q|\h\Phi\ra=-\beta\la H_\obs\ra$ in eqn. (\ref{urgo}) as $A_{\hor}/4G_N$ or more precisely as the shift
 in $A_{\hor}/4G_N$ due to the presence of the observer.

\section{Hilbert Spaces}\label{hilbertspaces}

\subsection{More On Type II Algebras and Hilbert Spaces}\label{more}

We have described an algebra $\A_P$ of observables that governs the experiences of an observer in the static patch $P$ of de Sitter space.  From the point of view
of an observer in $P$, the state of the universe is entirely summarized by a density matrix $\rho\in\A_P$.    
This observer has no way to know what there is outside of $P$.  

However, if we make a global model of the whole de Sitter space $X$, then we can construct a Hilbert space $\h\H$ that describes the whole universe. This
will give a purification of the density matrix $\rho$.    In general, a
state $\Psi\in\h\H$ governs observations both in the static patch $P$ and in the complementary static patch $P'$.    In this section, we will analyze the Hilbert
spaces that result from different assumptions about what is in the patch $P'$.

But first we will make a few general remarks about Hilbert space representations and density matrices.
As a preliminary, instead of the two static patches $P$ and $P'$, let us consider
ordinary quantum systems $A$ and $B$ with Hilbert spaces $\H_A$ and $\H_B$ that are respectively of dimension $n$ and $m$.   The combined system $AB$ has a tensor
product Hilbert space $\H_{AB}=\H_A\otimes \H_B$ of dimension $nm$.   If and only if $n=m$, any density matrix of either system is the reduced density matrix
of a pure state of the combined system.
  If $n>m$, then any density matrix of system $B$ can be realized by a pure
state of the combined system, but this is not true for system $A$.    For $n>m$, the maximum entropy of a density matrix of system $A$ that comes from a pure
state of the combined system is $\log m$.   This is less than the entropy $\log n$ of a maximally mixed state of system $A$ by an ``entropy deficit''  $\Delta S
=\log n-\log m=\log 1/d$, where $d=m/n$.   If instead $d>1$, the roles of systems $A$ and $B$ are reversed, and $B$ has an entropy deficit $\log d$.

All of this has a precise analog for an algebra $\A$ of Type II$_1$.   The role of $d$ is played by the ``continuous dimension'' of Murray and von Neumann,
which as explained in section \ref{whatis} classifies the representations of  $\A$ on a Hilbert space.  Here $d$ is a positive real number or infinity.
When $\A$ acts on a Hilbert space, its commutant
 $\A'$ 
is of Type II$_1$ unless $d=\infty$, in which case it is Type II$_\infty$.  

For $d=1$, we can assume that the Hilbert space $\H$ is a copy of the algebra $\A$ itself.   The inner product on $\H$ is defined by
$(\x,\y)=\Tr\, \x^\dagger\y$, for $\x,\y\in\H$.   The action of $\A$ on $\H$ is described by $\x\to \a\x$, for $\a\in\A$, $\x\in\H$.   The commutant of $\A$ is
another algebra $\A'$ that can be described as follows.  For any $\a\in\A$, there is an element $\a'\in\A'$, acting on $\H$ by right multiplication,
$\x\to \x\a'$.   ($\A'$ is actually the ``opposite algebra'' to $\A$, with multiplication in the opposite order: $(\a\b)'=\b'\a'$.)  
Let $\Psi\in\H$ be any pure state.   To find the reduced density matrix $\rho$ for the algebra $\A$, we compute
$(\Psi|\a|\Psi)=(\Psi,\a\Psi)=\Tr\,\Psi^\dagger \a\Psi=\Tr\,\a\Psi\Psi^\dagger$, where we used the cyclic property of the trace.   So $\rho=\Psi\Psi^\dagger$.
Similarly we can find the density matrix of the same state for the algebra $\A'$.   In this case, $(\Psi|\a'|\Psi)=(\Psi|\Psi\a')=\Tr\,\Psi^\dagger \Psi\a$
and therefore the density matrix is $\sigma=\Psi^\dagger\Psi$.

The formulas $\rho=\Psi\Psi^\dagger$, $\sigma=\Psi^\dagger\Psi$ show that by taking $\Psi=\rho^{1/2}$ or $\Psi=\sigma^{1/2}$, we can get an arbitrary density
matrix $\rho$ or $\sigma$ of algebra $\A$ or algebra $\A'$ as the reduced density matrix of a pure state $\Psi\in\H$.    
The same formulas also imply that $\Tr\,\rho^n=\Tr\,\sigma^n$, for any $n$, so the von Neumann and R\'{e}nyi entropies
of $\rho$ and $\sigma$ are always equal.   All this is as in ordinary quantum mechanics.

To get a Hilbert space representation of $\A$ with $d<1$, we can pick a projection operator $\Pi\in \A$ of trace $d$.   Then, viewing $\Pi$ an an element of $\A'$,
we project onto the subspace $\H\Pi\subset \H$ consisting of states $\Psi\Pi,$ $\Psi\in\H$.   The algebra $\A$ acts on the left on $\H\Pi$ as before.   Its
commutant consists of operators $\Pi\a'\Pi$, acting on the right by $\Psi\Pi\to \Psi\Pi\a'\Pi$.   These operators make an algebra $\t \A'=\Pi\A'\Pi$, which is also
of Type II$_1$.   The element $\Pi\in\A'$ is the identity in $\t\A'$.  It is usual in an algebra of Type II$_1$ to define a normalized trace such that the trace of the identity
element is 1. For the algebra $\t\A'$, the normalized trace is
\be\label{normtr}\t\Tr\,\Pi\a'\Pi=\frac{1}{\Tr\,\Pi}\Tr\,\Pi\a'\Pi. \ee   Clearly $\t\Tr\,\Pi=1$.

An example of a normalized state in $\H\Pi$ is $\Psi=\Pi/(\Tr\,\Pi)^{1/2}$.   Using the property $\Pi^2=\Pi$ of a projection operator,
one finds  that the density matrix of algebra $\A$ for this state is $\rho=\Psi\Psi^\dagger=\Pi/\Tr\,\Pi$.   The entropy of this density matrix is
\be\label{denred}S(\rho)=-\log (1/\Tr\,\Pi)=-\log (1/d), \ee
as we computed in section \ref{whatis}.   The normalized density matrix of the same state for the algebra $\t\A'$ is simply $\sigma=\Pi$, with entropy
\be\label{wenred}S(\sigma)=-\t\Tr\,\sigma\log\sigma=0. \ee
Thus, as in ordinary quantum mechanics, for $d<1$, there is a pure state of the system, namely $\Pi\in \H\Pi$, which has maximum entropy for the second
algebra $\t\A'$, but has an entropy deficit $\log(1/d)$ for  $\A$.

Now let us consider a general state $\Psi=\x\Pi\in \H\Pi$, with $\x\in\A$. 
For $\Psi$ to be normalized, we need
\be\label{normo}(\Psi,\Psi)= \Tr\, \x^\dagger\x \Pi =1.\ee
By similar calculations to those that we have already considered, the density matrix of the state $\Psi$ for the algebra $\A$ is
$\rho= \x\Pi\x^\dagger$, which satisfies $(\Psi|\a|\Psi)=\Tr\,\a\rho$, for all $\a\in\A$.   For $\Pi\a'\Pi\in\t\A'$, we get $(\Psi|\Pi\a'\Pi|\Psi)=\Tr\,\Pi\x^\dagger\x \Pi \a'
=\Tr\,\Pi\, \t\Tr \,\Pi \x^\dagger \x \Pi\a'$.   Therefore the density matrix of the same state for $\t\A'$ is
\be\label{wormo} \sigma=(\Tr\,\Pi) \Pi \x^\dagger\x \Pi.\ee
These formulas for $\rho$ and $\sigma$ lead to $\Tr\,\sigma^n=(\Tr\,\Pi)^n \Tr\,\rho^n$ for any $n$, and therefore $\t\Tr\,\sigma^n=(\Tr\,\Pi)^{n-1} \,\Tr\,\rho^n$.
Differentiating with respect to $n$ at $n=1$ to compute $S(\rho)=-\Tr\,\rho\log\rho$, $S(\sigma)=-\t\Tr\,\sigma\log\sigma$, we get
\be\label{enred} S(\rho) = S(\sigma) -\log (1/\Tr\, \Pi)=S(\sigma)-\log (1/d).   \ee
We already know that the state $\Psi=\Pi$ has the maximum entropy of any state in $\H\Pi$ for the algebra $\t\A'$, namely $0$, so this
formula shows that the same state has the maximum entropy of any state in $\H\Pi$ for the algebra $\A$, namely $-\log\,1/d$.  Thus, similarly to what happened
in ordinary quantum mechanics, the maximum entropy state in $\H\Pi$ has an entropy deficit $\log 1/d$ for the algebra $\A$.

As in ordinary quantum mechanics, exchanging the two algebras $\A$ and $\A'$ has the same effect as replacing $d$ with $1/d$, so we will not consider
separately the case $d>1$.

The case that is symmetrical between the two algebras is $d=1$.   This suggests that if we place in $P'$ an observer identical to the observer in $P$,
we might get a Hilbert space representation with $d=1$.  We will begin with this case, and show that it does lead to $d=1$.  As in the preceding discussion,
examples with $d\not=1$ can then be constructed by simply acting with a projection operator in one of the two algebras.  The case that there is no observer
in $P'$ turns out to be troublesome, and we will only be able to offer a conjecture about what happens in this case.

\subsection{An Observer in the Second Patch}\label{second}

In section \ref{constob}, we started with a Hilbert space $\H$ that describes quantum fields in de Sitter space, with the constraints ignored.  The important
constraint operator was the operator $H$ that generates future-directed time translations of the static patch $P$, and past-directed time translations of the complementary
patch $P'$.   Then we introduced an observer in $P$ with canonical
variables $p,q$, and a Hamiltonian $H_\obs=q$, with $q\geq 0$.        We  now extend this construction to  an identical observer in the complementary
static patch $P'$.    This observer has canonical variables $p',q'$ and Hamiltonian $H'_\obs=q'$, again with $q'\geq 0$.   The combined Hilbert space, ignoring
the Hamiltonian 
constraint, is therefore now $\H\otimes \H_\obs\otimes \H'_\obs$, where $p,q$ act on the Hilbert space $\H_\obs$ of the observer in $P$, and $p',q'$ act on the
Hilbert space $\H'_\obs$ of the observer in $P'$.   For example, we can represent $q$ and $q'$ as multiplication operators and set $p=-\i \partial/\partial q$,
$p'=-\i \partial/\partial q'$, or we can represent $p,p'$ by multiplication and $q,q'$ by differentiation.    We will, to begin with, ignore the conditions $q,q'\geq 0$,
and impose those conditions at the end by acting on the Hilbert space and the algebras with suitable projection operators.

The constraint operator of the combined system is the total Hamiltonian of the bulk system plus the two observers:
\be\label{totcon}\h H= H+H_\obs- H'_\obs=H+q-q'.\ee
The reason for the minus sign multiplying $H'_\obs$ is that $H$ generates past-directed time translations in the patch $P'$; indeed, $H$ is odd under the exchange
of the two patches, and the extended constraint operator $\h H$ has the same property.

We essentially already know from section \ref{constob} how to describe the algebras of the two patches.   The observer in patch $P$, if we ignore the constraint,
has access to an algebra $\A$ of operators on $\H$, and also to the operators $p,q$ on $\H_\obs$.   The combined algebra is $\A\otimes B(\H_\obs)$, where 
$B(\H_\obs)$ is the algebra of all bounded operators on $\H_\obs$.   Taking the constraint into account at the level of observables just means replacing $\A\otimes B(\H_\obs)$ with its $\h H$-invariant part.    Relative to section (\ref{constob}), the constraint now has an extra contribution $q'$, but since this operator commutes
with $\A\otimes B(\H_\obs)$, that makes no difference.   Hence the invariant subalgebra of $\A\otimes B(\H_\obs)$ is the same as it was before; it is generated by
$e^{\i p H}\a e^{-\i p H}$ for $\a\in \A$, along with $q$.   Schematically, the invariant algebra is
\be\label{welfo} \A^*_P= \{e^{\i p H}\a e^{-\i p H},q\},~~~\a\in \A .\ee  
The reason for the $*$ in $\A_P^*$ is that we have not yet imposed the conditions $q,q'\geq 0$; when those conditions are imposed, we will drop the $*$.
 The fact that $\A_P^*$ is not affected by the presence of an observer in the patch $P'$ is a special
case of the fact that an observer in patch $P$ does not know what is in $P'$.   
Similarly, before imposing the constraint, the algebra of the complementary patch $P'$ is $\A'\otimes B(\H'_\obs)$, where $\A'$ is the commutant of
$\A$ acting on $\H$.      
Imposing the constraint means replacing $\A'\otimes B(\H'_\obs)$ with its $\h H$-invariant part, which is generated by $e^{-\i p' H}\a' e^{\i p' H}$, $\a'\in
\A'$, along with $q'$.  Schematically
\be\label{zelfo} \A_{P'}^*   =\{ e^{-\i p' H}\a' e^{\i p' H},q'\}, ~~~\a'\in\A'.\ee
The algebras $\A_P^*$ and $\A_{P'}^*$ obviously commute.

However, we want to impose the constraint $\h H$ not just on the operators but on the Hilbert space  $\H\otimes \H_\obs\otimes \H'_\obs$.   It is straightforward to impose
a {\it compact} group $G$ of constraints on a Hilbert space: one just restricts to the $G$-invariant subspace of Hilbert space.   However, that does not work well for constraints that generate a noncompact group.  To understand why, let us consider a simplified case in which the constraint that we wish to impose is
just $q'$.    There is no Hilbert space of states annihilated by $q'$, since a state annihilated by $q'$ must be proportional to $\delta(q')$ and is not normalizable.
Our problem is not that different, because $-\h H=e^{-\i p'(H+q)} q' e^{\i p'(H+q)}$ is conjugate to $q'$.    In the case of a constraint $\h H$ that generates
a noncompact group (here the group of time translations of the static patch), rather than requiring a physical state $\Psi$ to satisfy $\h H\Psi=0$, it is often
better to impose an equivalence relation $\Psi\cong \Psi+\h H\chi$ 
for any $\chi$.   The equivalence classes are called coinvariants, and often one can define a Hilbert space
of coinvariants even though there is no Hilbert space of invariants.   The space of coinvariants has a natural interpretation in BRST quantization.  This is discussed
in detail in Appendix \ref{coinvariants}, but for the present case of a single constraint, we will give a brief explanation here.   In the case of a single constraint $\h H$,
the BRST complex is defined by introducing a single ghost operator $c$ and a single antighost operator $b$, satisfying $c^2=b^2=0$, $\{c,b\}=1$.   These
operators can be realized on a pair of states $|\negthinspace\downarrow\ra$, $|\negthinspace\uparrow\ra$, respectively of ghost number 0 and 1, satisfying
$|\negthinspace \uparrow\ra =c|\negthinspace\downarrow\ra$, $|\negthinspace\downarrow\ra=b|\negthinspace\uparrow\ra$.  The BRST operator is $Q=c\h H$.
The BRST cohomology is defined as the space of states $\h\Psi$ with $Q\h\Psi=0$ modulo the equivalence relation $\h\Psi\cong\h\Psi+Q\h\chi$.
The cohomology of ghost number 0 consists of states $\h\Psi=\Psi|\negthinspace\downarrow\ra$ annihilated by $Q$.   
The equivalence relation is vacuous for states of ghost number 0,  since
there are no states at ghost number $-1$.  The condition $Q\h\Psi=0$ is equivalent to $\h H\Psi=0$,  
so the BRST cohomology at ghost number 0 is the space of invariants.   
At ghost number 1, we have $\h\Psi=\Psi|\negthinspace\uparrow\ra$ for some $\Psi$. The constraint $Q\h\Psi=0$ is vacuous, 
since there are no states of ghost number 2, and the equivalence relation
$\h\Psi\cong \h\Psi+Q\h\chi$ becomes $\Psi\cong \Psi+\h H\chi$.   So the BRST cohomology at ghost number 1 is the space of coinvariants.  

The present problem is actually a typical example in which one wants to work with the space of coinvariants.    Represent $p'$ by 
multiplication, and $q'$ by  $q'=\i\partial/\partial p'$, and define a map from $\H\otimes \H_\obs \otimes \H'_\obs $ to $\H\otimes \H_\obs$ as follows.   
View an element $\Psi\in \H\otimes \H_\obs\otimes \H'_\obs$
as a function $\Psi(p')$ that is valued in $\H\otimes \H_\obs$.   For such a $\Psi$, define $T\Psi\in \H\otimes \H_\obs$ by
\be\label{deftop}T\Psi=\int_{-\infty}^\infty \d p' \,e^{\i p'(H+q)} \Psi(p'). \ee
Integration by parts shows that the map $\Psi\to T\Psi$ is invariant under $\Psi\to \Psi+\h H\chi$, with $\h H=H+q-\i \partial/\partial p'$.   
So $T$ gives a map from the space of coinvariants to $\H\otimes \H_\obs$, and this map is in fact an isomorphism.   So the space of coinvariants
can be identified with\footnote{This explanation is slightly oversimplified.   As in Appendix \ref{coinvariants}, the
 coinvariants have to be defined in a space of functions of $p$ with compact
support (or rapidly vanishing at infinity) rather than in a Hilbert space, and after defining the space of coinvariants, one then takes a completion to 
get a Hilbert space.   The result of a more careful analysis is as stated in the text.} $\H\otimes \H_\obs$, and this is the desired Hilbert space  $\h \H^*$ (on which we still have to impose $q,q'\geq 0$).   
Of course, we could have made a similar
construction with the role of the two observers exchanged, and then we would have identified $\h\H^*$ with $\H\otimes \H'_\obs$.

It remains to determine how the algebras $\A_P^*$ and $\A_{P'}^*$ act on $\h \H^*$.  By definition, any operator $\x$ in either $\A_P^*$ or $\A_{P'}^*$ commutes
with $\h H$ and therefore with the BRST operator $Q$; hence $\x$ makes sense as an operator on the BRST cohomology at ghost number 1, and therefore, on the space
of coinvariants.   Since we have identified the space of coinvariants with $\h\H^*=\H\otimes \H_\obs$ via the map $T$, there is a unique operator $\h \x$ on $\h\H^*$ such 
that $\h \x T\Psi =T(\x \Psi)$, and this is the operator by which $\x$ acts on $\h \H^*$.

When we carry out this procedure for $\A_P^*$, nothing happens.   $T$ commutes with the operators $p,q,H$, and $\a\in \A$ from which $\A_P^*$ is constructed.   So
as an algebra of operators on $\h\H^*$, $\A_P^*=\{e^{\i p H}\a e^{-\i p H},q\}$, $\a\in\A$,  exactly as in eqn. (\ref{zelfo}).    
What happens to $\A_{P'}^*$ is more interesting, since the definition of $\A_{P'}^*$ in eqn. (\ref{zelfo}) involves operators $q',p'$ that are eliminated when we identify
$\h\H$ as $\H\otimes \H_\obs$.
We find
\be\label{motlo}   T(e^{-\i p' H}\a' {e^{\i p' H}}\Psi)=\a' T\Psi\ee
and
\be\label{golfox} T(q'\Psi) = (H+q) T\Psi. \ee
So as an algebra of operators on $\h \H$, $\A_{P'}^*$ is generated by  $\a'\in \A'$ along with $H+q$:
\be\label{zolfox} \A_{P'}^*=\{\a',H+q\},~~~~~   \a'\in\A'. \ee
What we have arrived at in eqns. (\ref{zelfo}) and (\ref{zolfox}) is the usual picture of two commuting crossed product algebras acting on $\H\otimes L^2(\R)$,
with $q$ as a multiplication operator on $L^2(\R)$.  The same construction appears in analyzing a black hole \cite{GCP}.
We will reconsider the black hole in section \ref{bh}. 

As in section \ref{constob}, to put $\A_P^*$ in a standard form, we conjugate by $e^{-\i p H}$ and  define $x=-q$, leading to
\be\label{firstalg} \A_P^*=\{ \a, H+x\},~~~~~\a\in \A\ee
and
\be\label{secondalg}\A_{P'}^* =\{ e^{-\i p H}\a' e^{\i p H}, x\},~~~\a'\in\A'.\ee
Here
\be\label{welco}p=-\i\frac{\partial}{\partial q}=\i\frac{\partial}{\partial x}. \ee

We still have to impose the conditions $q,q'\geq 0$, which now become $H+x\leq 0 $ and $x\leq 0$. So we introduce projection operators 
$\Pi\in \A_P^*$, $\Pi'\in \A_{P'}^*$, defined by $\Pi=\Theta(-H-x), $   $\Pi'=\Theta(-x)$.   Then finally we define the physical Hilbert space 
in the presence of observers of nonnegative energy by projection
\be\label{proj} \h\H=\Pi\Pi'\h\H^*.\ee
   Similarly the physical algebras of the two static patches are
 \be\label{noj}
   \A_P=\Pi \A_P^*\Pi,~~~~\A_{P'}=\Pi'\A_{P'}^*\Pi'.\ee   The projection operators $\Pi\in\A_\P^*$, $\Pi'\in\A_{P'}^*$ become the identity operators
   in $\A_P$, $\A_{P'}$.

We already defined the  algebra $\A_P$  in section \ref{constob}. We showed it to be of Type II$_1$ and studied its normalized trace.  Since the construction is symmetric between the
two algebras, $\A_{P'}$ is also of Type II$_1$, and we expect that $\h\H$ has continuous dimension 1
 as a representation of $\A_P$ or of $\A_{P'}$.   This is equivalent to saying that the maximum entropy state
   of either algebra, with density matrix 1, can be realized by a pure state  $\Psi\in \h\H$.   The appropriate state, namely 
   $\Psi_{\mathrm{max}}=\Psi_\dS \,\sqrt{\beta_\dS}e^{\beta_\dS x/2}\Theta(-x)$,
   was already described in section    \ref{constob} (eqn. (\ref{pureds})).  The difference between the present discussion and the previous one is that
   in section (\ref{constob}), $\H\otimes L^2(\R_-)=\Pi(\H\otimes \H_\obs)$ was introduced arbitrarily as a Hilbert space in which the
   maximum entropy density matrix of the algebra $\A_P$ can be purified.   Now, instead, we have shown 
   that after a further projection by $\Pi'$, $\Pi(\H\otimes \H_\obs)$ becomes  the physical
   Hilbert space for the case that there are identical observers in the two patches.   
   The further projection by  $\Pi'$  does not affect the analysis of the density matrix of the state
   $\Psi_{\mathrm{max}}$ for the algebra $\A_P$,  since $\Pi'$ leaves $\Psi_{\mathrm{max}}$
   invariant and commutes with $\A_P$.
   
   As an interesting variant of this, we can consider the case that the complementary patch $P'$ contains an observer that is not isomorphic to the observer in $P$.
   As one simple option, we can assume that the Hamiltonian $H'_\obs=q'$ of the second observer is bounded in the range $a\leq q'\leq b$.   We do not have to 
   assume that $a\geq 0$, but it is physically natural to do so.  Moreover, in that case, the projection operator onto states with $a\leq q'\leq b$, namely 
   $\Pi'_{[a,b]}=\Theta(b-q')\Theta(q'-a)=\Theta(b+x)\Theta(-x-a)$, is an element of the algebras $\A_{P'}$.   The computation of $\Tr\,\Pi'_{[a,b]}$ 
   proceeds precisely as in eqn. (\ref{comdel}),
   except that the integral over $x$ goes over the range $-b\leq x\leq -a$:
   \be\label{newtrace}\Tr\,\Pi'_{[a,b]} =\int_{-b}^{-a}\beta_\dS \d x \,e^{\beta_\dS x} = e^{-\beta_\dS a}-e^{-\beta_\dS b}. \ee
   This illustrates the assertion that a projection operator in a Type II$_1$ algebra can have any trace between 0 and 1, as discussed in section \ref{whatis}.

   In the presence of an observer in the patch $P'$ with energy bounded in the range $a\leq q'\leq b$, the Hilbert space becomes
   $\h \H_{[a,b]}=\Pi \,\Pi'_{[a,b]} \h\H^*$.   The algebra $\A_P$ is unaffected, but we now have $\A_{P'}=\Pi'_{[a,b]}\A_{P'}^* \Pi'_{[a,b]}$.    As a representation of $\A_P$,
   $\h\H$ now has continuous dimension $d=\Tr\,\Pi'_{[a,b]}$.  As explained in section \ref{more}, the maximum possible entropy of a pure state in $\h\H_{[a,b]}$, given
   that this Hilbert space has continuous dimension $d<1$ for the algebra $\A_P$, is 
   \be\label{smax} S_{P,\mathrm {max}}= -\log (1/d)\ee
    for the algebra $\A_P$, and $S_{P',\mathrm{max}}= 0$  for $\A_{P'}$.
     A normalized state that has the maximum possible entropy for each algebra is
   \be\label{newmax}\Psi_{\mathrm{max},[a,b]}=\frac{1}{\sqrt{\Tr\,\Pi'_{[a,b]}} }\Pi'_{[a,b]} \Psi_{\mathrm{max}}. \ee
   
   The physical meaning of the entropy reduction in the patch $P$ is that as we have removed some qubits from the patch $P'$, there is no state in $\h\H_{[a,b]}$
   that describes complete entanglement of all degrees of freedom in $P$ with anything in $P'$.  On the other hand, the the state $\Psi_{\mathrm{max},[a,b]}$ describes complete entanglement of everything in $P'$ with something in $P$.

   \subsection{No Observer in the Second Patch}\label{emptypatch}
   
   If we try to repeat this analysis without assuming the presence of an observer in the static patch, we run into immediate difficulty.
   
   We found in section \ref{second} that, assuming there is an observer in the complementary patch $P'$, we can incorporate the constraints at the level of
   Hilbert space states by simply omitting the Hilbert space $\H'_\obs$ of this second observer.  Whether there is an observer in the patch $P$ was not important in this
   argument.   In other words, if one assumes that there is an observer in $P'$ but none in $P$, then the map $T$ defined in eqn. (\ref{deftop}) makes sense
   as an isomorphism between the space of coinvariants in $\H\otimes \H'_\obs$   and $\H$.    So in this case, the physical Hilbert space on which $\A_{P'}^*$ acts
   should be just $\h\H=\H$.
   
   Since the definition of the map $T$ would make sense with the roles of $P$ and $P'$ exchanged, we also seem to learn that if there is an observer in $P$ and
   none in $P'$,  the algebra $\A_P^*$ should act on $\h\H=\H$. 
However, this is not the case. Recall from eqn. (\ref{zolfox}) that, as an algebra of operators on $\h\H$, $\A^*_{P'}$ is generated by $\a' \in \A'$ and $H+q$. For the same reason, the action of $\A^*_P$ on $\H$ should now be generated by $\a \in \A$ and $H$. As we established earlier, for any representation $\h\H$ of $\A^*_P$, the commutant algebra ${\A^*_P}'$ should be a Type II von Neumann algebra. But the reason that we introduced an observer in the patch $P$ in 
the first place is that on the Hilbert space $\H$ of quantum fields in de Sitter space, there are no operators (except $c$-numbers) that commute with both $\A$ and $H$. So $\H$ cannot be a representation of $\A^*_P$.

We do not have a definitive resolution of this puzzle, but we will offer
a conjecture.

To motivate the conjecture, we start with this question.    How much less do we expect the maximum entropy to be if there an observer in $P$ but none in $P'$,
relative to the case of identical operators in each of the two patches?

To motivate an answer, let us discuss the corrections to the entropy $S_\dS$ of de Sitter space.  According to Gibbons and Hawking \cite{GH}, in leading order  $S_\dS=A/4G_N$, where $A$ is the area of the cosmological horizon.   They obtained this result by evaluating the path integral for the Euclidean version of
de Sitter space, which is simply a sphere.   In the approximation of keeping only factors that depend exponentially on $G_N$, the result is $e^{-I}$,
where $I=-A/4G_N$ is the classical action.   Gibbons and Hawking interpreted this as $e^{S_\dS}$, giving $S_\dS=A/4G_N$.

A more precise evaluation of the Euclidean path integral would then be expected to give a better approximation to $S_\dS$.   For our purposes, what is 
relevant is an extremely simple correction\footnote{The closest analog of this correction for a black hole would appear in the normalization of the measure
for integration over the time-shift mode.}  to the original calculation that shifts the entropy by a multiple of $\log G_N$.   In general, consider a  gauge or gravitational
theory with a classical action 
proportional to $1/\lambda$, where $\lambda$ is a coupling parameter.
In such a theory, consider a family 
 of classical solutions that depends on $r$ parameters,
and assume that the generic element of the family is invariant under an $s$-dimensional group $F$ of gauge symmetries.   Then the
contribution of this family to the path integral is proportional to $\lambda^{-\frac{1}{2}(r-s)}$ (times $e^{-I}$, with $I$ the action).
In the case of gravity, we can use this formula with $\lambda=G_N$.   As a solution of Einstein's equations with positive cosmological constant, de Sitter space
does not depend on any moduli, so this is a case with $r=0$.   On the other hand, the de Sitter group $G_\dS=\SO(D+1)$ is a group of symmetries of
this solution, so $s=\dim\,G_\dS$.   So a slightly more precise formula for the Euclidean path integral of de Sitter space is $e^{-I} G_N^{\dim\,G_\dS/2}.$
Equating this with $e^S$, one finds a logarithmic correction to the de Sitter entropy
\be\label{jumper}S_\dS=\frac{1}{4G_N} -\frac{\dim\,G_\dS}{2}\log(1/G_N)+\cdots, \ee
where the omitted terms are of order 1 for $G_N\to 0$.    This logarithmic correction can be found in eqn. (1.4) of \cite{Denef} and more explicitly (for some values of $D$)
in eqn. (1.12) of that paper.

Now assume that a particular static patch $P$ has been selected by some sort of boundary condition or partial gauge-fixing (see section \ref{ea})
that explicitly breaks $G_\dS$ down to the symmetry
group  of $P$.
  This group is $G_P=\R\times \SO(D-1)$, where $\R$ is the group of time translations of $P$, and $\SO(D-1)$ is the group
of rotations.  So the logarithmic correction to the entropy is now $-\frac{1}{2}\left(1+\dim \,\SO(D-1)\right) \log(1/G_N)$.   

What will happen to the maximum possible entropy 
 if we include an observer in the patch $P$, and nothing in $P'$?  In this case, we expect no increase in the
maximum possible
entropy, since in the maximally entropic state of de Sitter space, all modes in $P'$ are  fully entangled.  Adding something to $P$ without adding anything to $P'$ does
not increase the maximum possible entanglement entropy, which is limited by what there is in $P'$.   
  So we expect that if an observer is added in $P$ only, the logarithmic correction to the entropy
is unchanged and remains $-\frac{1}{2}(1+\dim\,\SO(D-1))\log(1/G_N)$.    On the other hand, suppose we add an observer in $P'$ who is identical to the observer in $P$.   
Then the two observers can be maximally entangled.  A calculation similar to the Gibbons-Hawking calculation but including the entangled pair of observers
is discussed in section \ref{ea}.   Intuitively, without getting into details,
 we should expect that with the observers present, time translations should not be interpreted as a group of unbroken
gauge transformations, because time translations act on the degrees of freedom of the observers.   However, unless we equip each observer with an 
orthonormal frame (this option was discussed in section \ref{frame}), rotations are still a group of gauge symmetries.   So we expect that the logarithmic
correction in the presence of the pair of entangled observers will be just $-\frac{1}{2} \dim\,\SO(D-1)\log (1/G_N)$.
In other words,  omitting an observer from the patch $P'$ reduces the maximum entropy by $\frac{1}{2}\log (1/G_N),$ up to a remainder that is finite for $G_N\to 0$.

With this in mind, let us go back to the question of what is the appropriate Hilbert space $\h\H$ if there is an observer in the patch $P$ and none in the patch
$P'$.   This Hilbert space is expected to admit an action of the Type II$_1$ algebra $\A_P$.   It therefore can be characterized by a continuous dimension $d$,
and, if $d<1$,  the maximum entropy state of $\A_P$ that can be realized by a pure state in $\h\H$ has an entropy deficit $\log (1/d)$, as we learned in section \ref{more}.
Since the entropy deficit that we expect as a result of having no observer in $P'$ is $\frac{1}{2}\log (1/G_N), $ we are led to conjecture that
\be\label{tofffo} d\sim G_N^{1/2} .\ee

If this is the correct answer, then it is not surprising that we do not get a sensible answer when we apply the analysis of section \ref{second} to the case of an observer
in only one patch, because our analysis only suffices to determine the limit of $d$ for $G_N\to 0$, and eqn. (\ref{tofffo}) says that this limit is the forbidden value
$d=0$.     If eqn. (\ref{tofffo}) is correct, some sort of quantum correction is needed in order to get a sensible result.
Possibly it is important to take into account fluctuations in the
horizon.

\subsection{Euclidean Approach}\label{ea}

Here we will generalize the Euclidean approach to the de Sitter entropy \cite{GH} to include a completely entangled pair of observers.    The Euclidean approach as traditionally developed
computes a function that is called the ``entropy'' without including an observer; however, it is entirely unclear in what sense the function that is computed actually is an entropy.
By including an observer in the Euclidean description, we can describe a calculation that is similar in spirit to the original one \cite{GH} and that gives a similar answer (with a different
coefficient for the logarithmic correction), but does have a clear interpretation in terms of entropy.

  With our assumption
that an observer is described by operators $p,q$ and Hamiltonian $H=q$, the observer can be described in Lorentz signature by an action $\int \d\tau (p\dot q - q)$,
where $\tau$ is proper time measured along the observer's worldline.  In Euclidean signature, this becomes $\int\d\tau (-\i p \dot q - q)$.

 \begin{figure}
 \begin{center}
   \includegraphics[width=1.75in]{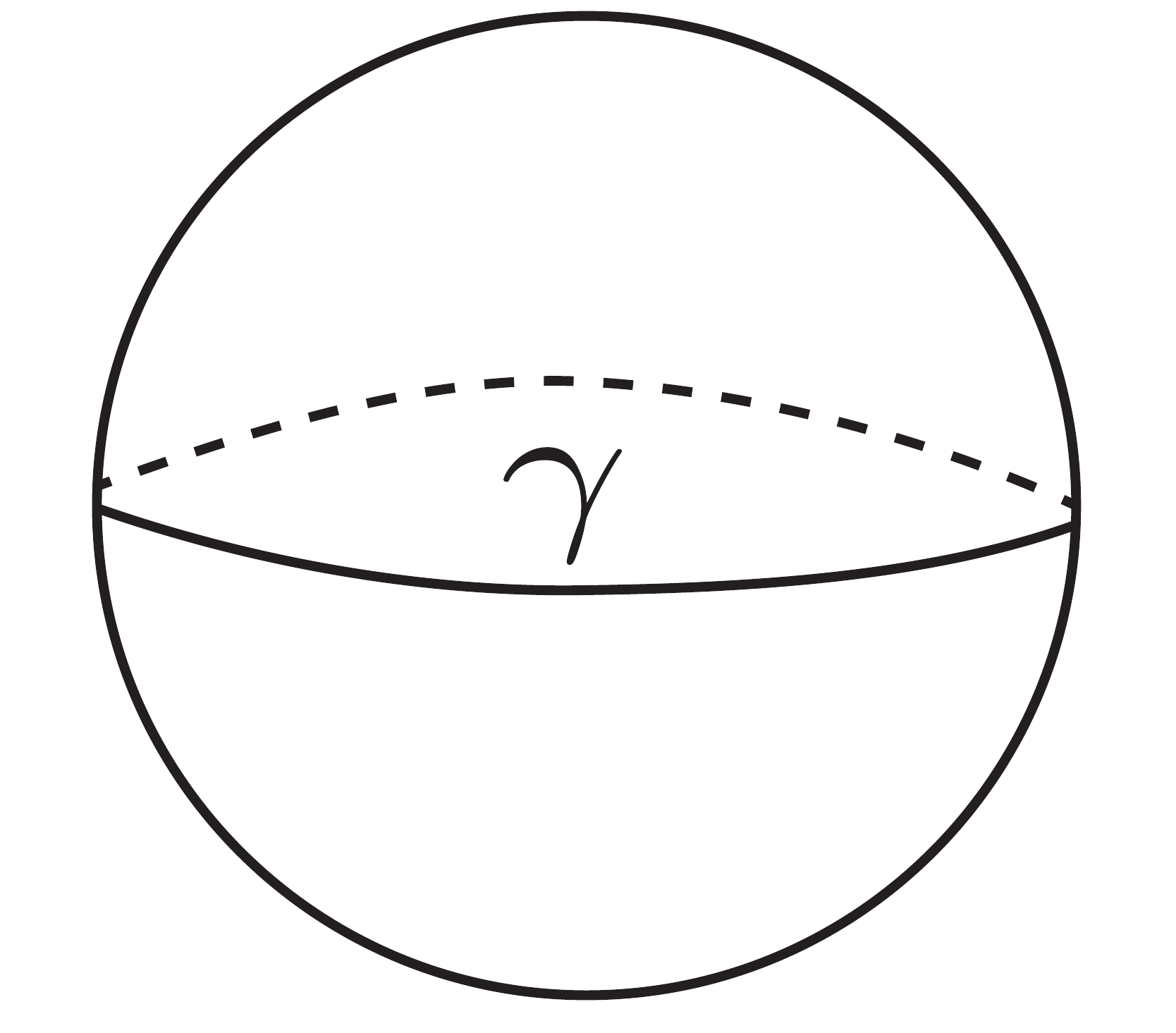}
 \end{center}
\caption{\small A sphere $S^D$ -- drawn here for $D=2$ -- with a closed geodesic $\gamma$, the Euclidean worldline of the observer.  \label{BBB}}
\end{figure} 

The Euclidean version of  de Sitter space is a sphere $S^D$.  We assume that the path integral for de Sitter space containing an entangled pair of observers
is dominated by a saddle point in which spacetime is $S^D$ with a round metric, and the observer worldline is a simple closed geodesic $\gamma\subset S^D$ 
(fig. \ref{BBB}).
The circumference of $\gamma$ is $\beta_\dS$, the inverse of the de Sitter temperature.   

In this situation, if
we ``cut'' $S^D$ on a plane of symmetry that is orthogonal to the worldline $\gamma$, and continue to Lorentz signature, then we get a Lorentz signature
picture with a pair of observers, one in a static patch $P$ and one in the complementary patch $P'$.   We expect these two observers to be maximally entangled
in a thermofield double state, since in the Euclidean picture they are connected by a finite length worldline.

In Lorentz signature, a particular static patch $P$ can be selected by imposing a boundary condition on
the observer's worldline at past and future infinity.   In Euclidean signature, there is no asymptotic region
in which a boundary condition can be imposed.   Instead, we select a particular closed geodesic  $\gamma\subset 
S^D$ by partial gauge-fixing, reducing the symmetry group $G_\dS$
of   de Sitter space to the symmetry group of a particular geodesic, which is the same as the symmetry group $G_P$ of a particular static patch.    Taking this into account,
 the coefficient of the logarithmic correction discussed in section \ref{emptypatch} is reduced from $-\frac{1}{2}\dim\,G_\dS$ to $-\frac{1}{2}\dim \,G_P$.

Apart from this, in the leading approximation for small $G_N$, the path integral that we have to do factors as the product of a gravitational path integral,
studied originally by Gibbons and Hawking \cite{GH} and in much more detail recently \cite{Denef}, and a path integral for the observer.

The path integral of the observer is 
\be\label{tombo}\int Dp(\tau)\,Dq(\tau)\,\exp\left(\oint_\gamma \d\tau \left(\i p\frac{\d q}{\d \tau} - q\right)\right). \ee
We do first the  path integral over $p$.  This gives a delta function $\delta(\d q/\d \tau)$.   This delta function has two consequences.

First, in doing the path integral over $q$, we can treat $q$ as a constant.    The action for the case that $q$ is constant is $\beta_\dS q$, so the measure in the integration
over the zero-mode of $q$ is $e^{-\beta_\dS q}$.   Bearing in mind that in analyzing the algebra  in section (\ref{constob}) we set $x=-q$, this factor is $e^{\beta_\dS x}$.   We have seen
this factor  before.   The pure state with maximum entropy for the algebra $\A_P$ of the static patch $P$
was $\Psi_\max =\sqrt{\beta_\dS} e^{\beta_\dS x/2}\Psi_\dS$.   In a Type I context, such a state would lead to a density matrix
proportional to  $e^{\beta_\dS x}$, although in the usual 
formalism for
Type II algebras, this  exponential factor is in the definition of the trace  (eqn. (\ref{tradef})) rather than in the density matrix (which is $\rho=1$ for the 
maximally entropic state).

But the delta function has another and more troublesome consequence.  For any periodic function $q(\tau)$, we have $\oint_\gamma\d \tau \frac{\d q}{\d \tau} =0$.
Therefore the delta function $\delta(\d q/\d\tau)$ that came from the integral over $p$ contains a factor $\delta(0)$.   
Concretely, this factor comes from the integral over the constant mode of $p$.
Formally an infinite factor $\delta(0)$ in the path integral  
 increases the logarithm of the path integral, which is  the de Sitter entropy, by an infinite additive constant.

We argued in section \ref{emptypatch} that adding an entangled pair of observers in the patches $P$ and $P'$ increases the maximum possible entropy not
by an infinite amount but by an amount of order $\frac{1}{2}\log\, G_N$.   To get this answer, we would like to replace the factor $\delta(0)$ that came from
the path integral over $p$ by a factor $1/G_N^{1/2}$.     How can we do this?

By taking the Hamiltonian of the observer to be simply $H=q$, we have given a continuous spectrum to the observer and therefore it is natural that incorporating a pair
of entangled observers can increase the maximum entanglement entropy by an infinite amount, accounting for the $\delta(0)$.   Thus from this point of view,
to replace $\delta(0)$ by $1/G_N^{1/2}$, we should replace the continuous spectrum of the observer by a discrete spectrum with a level spacing of order $G_N^{1/2}$.

A simple way to do this is to replace the Hamitonian $H=q$ of the observer by $H=G_N p^2+q$, which does indeed lead to a level spacing of order $G_N^{1/2}$.
If we add this term to the Hamiltonian, then the action likewise has an extra term $G_N\oint_\gamma \d \tau\, p^2(\tau)$.   For small $G_N$, this has no important
effect in the preceding calculation except to replace the $\delta(0)$ that came from the integral over the constant mode of $p$ with a factor $1/G_N^{1/2}$.
Moreover, it seems that a $G_N p^2$ term in the Hamiltonian of the observer would have played little role elsewhere in this article, since all of our considerations
have involved the limit $G_N\to 0$.

Although it may be true that adding a term $G_N p^2$ to the Hamiltonian of the observer makes a better model, there are two things that make us reluctant to propose
this as a solution to the problem with the $\delta(0)$.   First, it is not clear to us on physical grounds why one should assume that the observer has a level spacing
of order $G_N^{1/2}$.   Second, it is not clear that adding this term to the Hamiltonian of an observer would help in addressing the possibly related puzzle
that we ran into in section \ref{emptypatch}, where it was difficult to understand the continuous dimensions of the 
physical Hilbert space in the case that there is an observer in only one of the
two complementary patches.    Something more subtle may be needed to deal properly with both the $\delta(0)$ and the continuous dimension.

\section{Revisiting the Black Hole}\label{bh}

 In this section, we will revisit the Type II$_\infty$ algebra of observables exterior to a black hole horizon.  One goal is to simplify the arguments that have been  presented
previously, and formulate them in a way that is not limited to the case of negative cosmological constant.   A second goal is to explain that the black hole can be treated quite similarly to the way that we have analyzed de Sitter space in the present article.

For  illustration, we will primarily focus on the maximally extended Schwarzschild solution in an asymptotically flat spacetime. 
However, our arguments apply equally well for AdS Schwarzschild, and we will briefly turn to this setting to discuss the boundary 
interpretation of the Type II$_\infty$ algebra. 

We work in a semiclassical limit in which the Schwarzschild radius $r_S$ of the black hole is held fixed, 
while Newton's constant $G_N \to 0$. This means that the mass of the black hole
is very large.

\subsection{The Type II$_\infty$ algebra}\label{type2}

 \begin{figure}
 \begin{center}
   \includegraphics[width=3in]{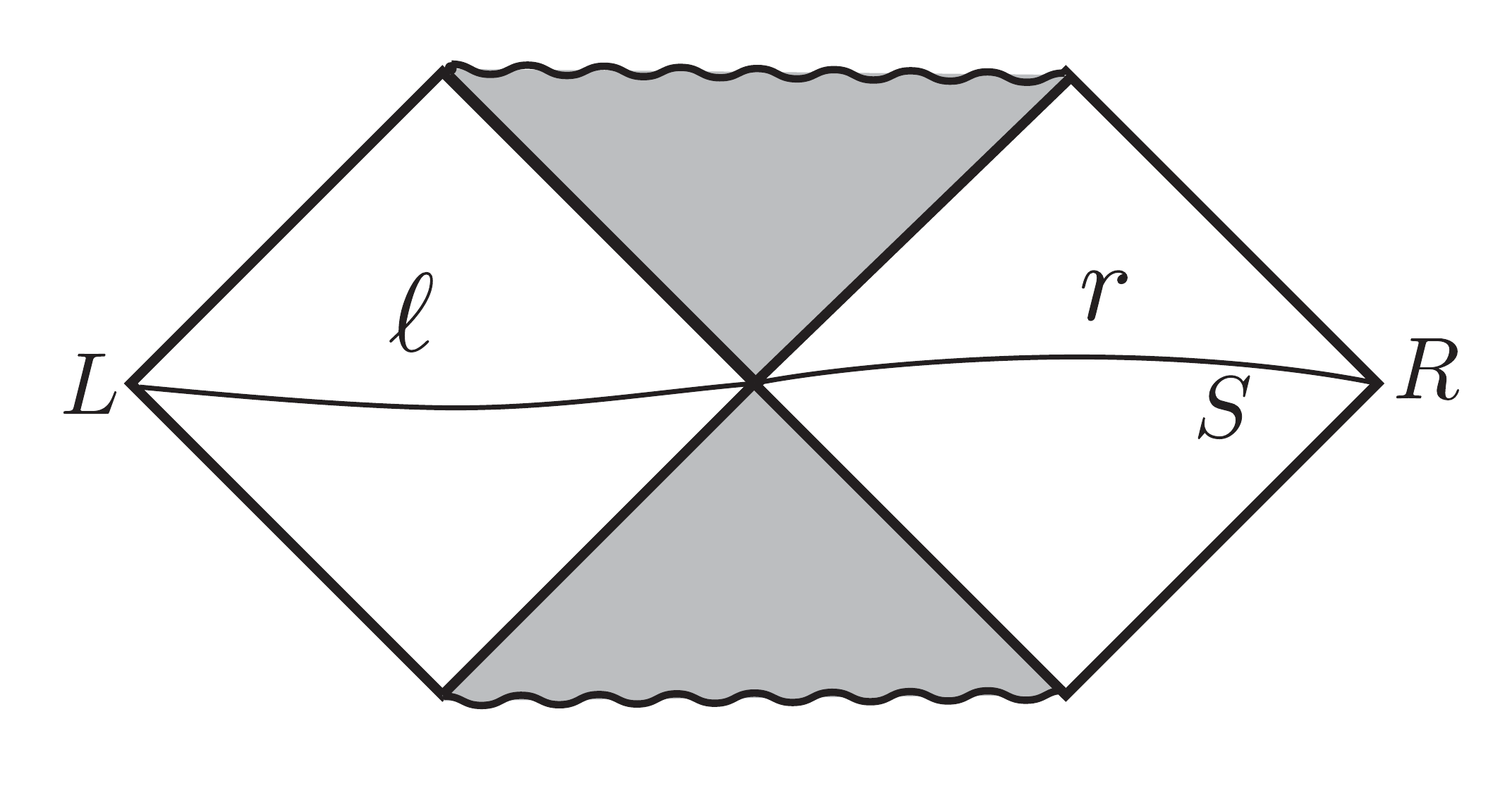}
 \end{center}
\caption{\small The Penrose diagram of the maximally extended asymptotically flat Schwarzschild spacetime.   The left and right exteriors are the unshaded regions
labeled
$\ell$ and $r$; the left and right regions at spatial infinity are labeled  $L$ and $R$.   The shaded regions are behind past and future horizons.  $S$ is
a bulk Cauchy hypersurface, chosen so that it can be decomposed as a union of portions in regions $\ell$ and $r$.  \label{CCC}}
\end{figure} 

Let $Y$ be the maximally extended Schwarzschild spacetime (fig. \ref{CCC}), and let $\ell$ and $r$ denote the left and right exteriors in $Y$ ($L$ and $R$ will
denote left and right spatial infinity).   
Consider first ordinary quantum field theory in the curved spacetime $Y$. This theory has a Hilbert space $\mathcal{H}_0$, and Type III$_1$ 
algebras $\A_{\ell,0}$ and $\A_{r,0}$ of observables in the left and right exterior.  The algebras $\A_{\ell,0}$ and $\A_{r,0}$ are factors (their centers
consist only of complex scalars), and they are each other's commutants.

The spacetime $Y$  has a Killing vector field $V$
that generates time translations.   $V$ is future-directed in the right exterior and past-directed
in the left exterior.   In the quantum theory constructed on $Y$ by expanding around the black hole
solution, there is a conserved quantity $\h h$ associated to $V$.  If $S$ is a bulk Cauchy hypersurface, then 
\be\label{concharge} \h h =\int_S\d \Sigma^\mu V^\nu  T_{\mu\nu}, \ee
where $T_{\mu\nu}$ is the energy-momentum tensor of the bulk fields (including
the energy-momentum pseudo-tensor of the bulk gravitational fluctuations).
The operator $\h h$ has an interpretation in modular theory (or Tomita-Takesaki theory) that is related to the thermal nature of the black hole.    
Let  $\Psi_{HH}$ be the Hartle-Hawking state of the quantum fields on $Y$ \cite{HH,I}. Then the modular operator
of the state $\Psi_{HH}$ for the algebra $\A_{r,0}$ is $\Delta=\exp(- H_\mod )$ where $H_\mod$, known as the modular Hamiltonian, satisfies  $H_\mod=\beta \h h$, 
with $\beta$  the
inverse of the Hawking temperature \cite{BW,Sewell}.

As in fig \ref{CCC}, one  can choose the hypersurface $S$  to pass through the codimension 2 intersection of the left and right horizons.
It can then be split as a union $S=S_\ell \cup S_r$ of portions to the left and right of the horizon.   
Having done so, it is tempting to define left and right bulk Hamiltonians\footnote{The reason to define $h_\ell$ with a minus sign is that $V$ is past-directed in the left
region.}
\be\notag h_r  =  \int_{S_r} \d \Sigma^\mu V^\nu  T_{\mu\nu}  \ee
 \be\label{zelf}  h_\ell = - \int_{S_\ell} \d \Sigma^\mu V^\nu  T_{\mu\nu} ,\ee
   and then
   \be\label{wofo} \h h = h_r - h_\ell. \ee
   However, this does not work, because the fluctuations in $h_r$ and $h_\ell$ are divergent. 
   A modular Hamiltonian of a Type III algebra never has a splitting as in eqn. (\ref{wofo}).  By contrast, for algebras of 
   Type I or Type II, there is always such a splitting. As we shall see, adding gravity will lead to a splitting of $\h h$ and  to a Type II$_\infty$ algebra.

In the limit $G_N \to 0$, local fluctuations of the gravitational field can be treated as just one more field propagating on $Y$. 
Specifically, for $G_N \to 0$, the graviton becomes a massless spin 2 free field, which can be quantized in the usual way. 
However, adding gravity also introduces additional modes that are not related to local excitations. 
This is perhaps most obvious in the case of JT gravity in two dimensions.    There are no local excitations at all.   Nevertheless, JT gravity has a two-dimensional
phase space, which was analyzed in detail in \cite{JaffHar}, for example.  The phase space consists by definition of modifications of the geometry along a
Cauchy hypersurface $S$ that satisfy natural boundary conditions at infinity, 
modulo those that can be eliminated by a diffeomorphism of the full spacetime that is trivial at infinity.
One such mode is the black hole energy, which in JT gravity is the same whether
it is measured in the left or right asymptotic region.   In other words, in JT gravity, the left and right Hamiltonians are equal, $H_L = H_R$.    The second mode 
in JT gravity is more subtle; it involves a shift  in the times $t_L$ and $t_R$ measured at infinity along $S$ on the left and on the right.   In other words, 
a localized perturbation of the geometry along $S$ can have the property that it can be eliminated by a diffeomorphism, but only by a diffeomorphism
that acts nontrivially at infinity, shifting the times $t_L$ and $t_R$ measured along $S$ in the left and right asymptotic regions.
Since the Schwarzschild spacetime has a time translation symmetry that shifts
$t_L$ and $t_R$ by equal and opposite amounts, only the sum $\Delta = t_L +  t_R$ is well-defined; thus there is only one mode of this kind.

   The time-shift  $\Delta$ is not measureable
by an observer on either the left or right side of the black hole, so it is not part of the algebra of observables in either exterior region.   In JT gravity, the algebras
of gauge-invariant observables accessible on the left or right regions outside the horizon are generated, respectively, by $H_L$ and $H_R$.    In particular,
these algebras are abelian, and the two algebras coincide, since $H_L=H_R$.   This is quite different from the situation in ordinary quantum field theory,
where the left and right algebras $\A_{\ell,0}$ and $\A_{r,0}$ are factors and their intersection consists only of multiples of the identity.

For Schwarzschild black holes above two dimensions, the  time-shift mode is still present, and there are additional modes of a somewhat similar
nature that represent a relative rotation between the two asymptotic boundaries.   
  Taking these modes into account, the full  Hilbert space in a Schwarzschild spacetime, 
including perturbative gravitational fluctuations  around the Schwarzschild
background,  is
\be \label{eq:Hphys}
\mathcal{H} \cong \mathcal{H}_0 \otimes L^2(\R_t) \otimes L^2(\SO(D-1)),
\ee
where $L^2(\R_t)$ and $L^2(SO(D-1))$ are respectively the spaces of square-integrable functions of the  time-shift $\Delta$ and of the relative rotation.   
We give an alternative derivation of this result -- making clear the close analogy with the Hilbert space 
of two observers in de Sitter space -- in Section \ref{sec:desitteranalog} below.     Of the three factors in $\H$, the first, namely $\H_0$, by itself, admits the action of
natural Type III algebras in the left and right exterior regions, just as in ordinary quantum field theory in a Schwarzschild background.   Our main goal here is to explain that properly including the
second factor $L^2(\R_t)$ turns the Type III algebras into algebras of Type II$_\infty$.    The third factor
$L^2(\SO(D-1))$ describes the possibility that the Schwarzschild black hole can acquire spin and become a  Kerr black hole.     (To be more precise, a
 slowly rotating black hole
can be described by a wavefunction in $L^2(\SO(D-1))$, but if the black hole rotation is not sufficiently slow, one must instead use the nonlinear Einstein
equations and the full Kerr solution.)        This third factor does not affect the
nature of the algebra, and so we will henceforth ignore it for simplicity.

To analyze the algebra of observables in the right exterior region, it is convenient to use the time translation symmetry of the Schwarzschild solution
to fix $t_R=0$ and therefore $\Delta= t_L$.   Having made this choice, the coordinate system at spatial infinity in the right exterior is completely fixed.   Therefore, it is possible to gravitationally dress bulk operators
to the right asymptotic region.   To gravitationally dress bulk operators to the left exterior region,   we would instead choose
$ t_L=0$, $\Delta = t_R$.

Time translations of the left and right boundaries are generated by the ADM Hamiltonians $H_L$ and $H_R$.   We aim to construct an algebra of
observables for the right exterior region in the limit $G_N\to 0$.   In that limit, $H_L$ and $H_R$ diverge, simply because the black hole mass $M_0=r_S/2G_N$ 
diverges.   It is convenient, therefore, to define subtracted Hamiltonians
\be\label{redham} h_L=H_L-M_0,~~~~~h_R=H_R-M_0 \ee
that have limits for $G_N\to 0$.    

The operator $h_L$ generates shifts of $t_L$.    We can make a  unitary transformation to reduce to the case that
$h_L$ acts only on $t_L$, $h_L=\i \frac{\partial}{\partial t_L}$, since if instead $h_L=\i\frac{\partial}{\partial t_L}+X_L$ where $X_L$ acts
on $\H_0$, we can remove $X_L$  by conjugating with $\exp(\i t_L X_L)$, and thereby reduce to $h_L=\i \frac{\partial}{\partial t_L}$.

The  conserved charge $\h h$ associated to the time translation symmetry of the full spacetime is related to the right and left ADM Hamiltonians by
\be \label{ADMconstraint}
\h h=H_R-H_L= h_R-h_L.
\ee
The algebra $\A_r$ of operators in the right exterior is generated by the algebra $\A_{r,0}$ of QFT operators in the right exterior, together with the subtracted right ADM energy $h_R = h_L + \h h$. Since $h_L$ commutes with everything in  $\A_{r,0}$, while $\h h$ generates the modular automorphism group of $\A_{r,0}$,
this is the crossed product of $\A_{r,0}$ by the modular automorphism group of $\Psi_{HH}$, and hence is a Type II$_\infty$ algebra.

The algebra $\A_\ell$ of operators in the left exterior is generated by the subtracted left ADM energy $h_L$, together with $e^{\i t_L \h h}\A_{\ell,0} e^{-\i t_L \h h}$. The conjugation of the left exterior QFT algebra $\A_{\ell,0}$ by $e^{\i t_L \h h}$ is necessary because we chose a gauge where operators that act solely on $\mathcal{H}_0$ are implicitly dressed to the right boundary, and hence are not local to the left exterior. Conjugating by $e^{\i t_L \h h}$ shifts this gravitational dressing to the left boundary. Explicitly, one can check that
\be
\left[ e^{\i t_L \h h}\A_{\ell,0} e^{-\i t_L \h h}, h_L + \h h \right] = 0~,
\ee
and hence $ e^{\i t_L \h h}\A_{\ell,0} e^{-\i t_L \h h}$ commutes with right-boundary time translations. Indeed, the algebra $\A_\ell$ is the commutant of $\A_r$, as desired.

\subsection{Isometries and Gauge Constraints} \label{sec:desitteranalog}

 We now give an alternative derivation of the Hilbert space and algebra described above that makes the analogy with the de Sitter construction manifest. The isometry group of the asymptotic boundary of the Schwarzschild spacetime $Y$ includes two copies of $\R_t$ that act separately as time translations of the left and right boundaries.\footnote{For simplicity, we omit two copies of $SO(D-1)$ whose inclusion would lead to the relative orientation mode in \eqref{eq:Hphys}. We also omit certain 
 asymptotic symmetries such as translations that do not interact with the gauge constraints in an interesting way.}  Gauge transformations of the asymptotic boundary are physical degrees of freedom. Hence, before imposing any gauge constraints, we have an ``extended Hilbert space''
$$
\mathcal{H}^{\mathrm{ext}} \cong \mathcal{H}_0 \otimes L^2(\R_t)_L  \otimes L^2(\R_t)_R~.
$$
Here, $\mathcal{H}_0$ describes the state of the bulk quantum fields, $L^2(\R_t)_L$ describes timeshifts of the left asymptotic boundary, and $L^2(\R_t)_R $ describes the timeshifts of the right asymptotic boundary. Just as in section \ref{type2},
up to conjugation, the  left and right subtracted ADM Hamiltonians act solely on $L^2(\R_t)_L$ and $L^2(\R_t)_R$ as $h_L = \i \partial_{t_L}$ and $h_R = \i \partial_{t_R}$ respectively. The (extended) right exterior algebra consists of 
\be \label{eq:extendedalgebra}
\A_r^{\mathrm{ext}} \cong \A_{r,0} \otimes B\left(L^2(\R_t)_R\right),
\ee
with the left exterior algebra defined analogously. 

Note that these are almost the same algebras and Hilbert space that we found for de Sitter space with two observers  in section \ref{second}, before imposing the gauge constraints. The only difference is that the subtracted ADM energies $h_L$ and $h_R$, unlike the observer energies, do not need to be positive.\footnote{In fact, as we shall see, the analogous constraint to the positivity of the observer energies in de Sitter would be a finite {\it upper} bound on the subtracted ADM energies $h_L$ and $h_R$.} Instead,  $h_L$ and $h_R$ can take any real values in the semiclassical limit, because of the large constant $r_S/2G_N$ that has been subtracted when defining them. As in section \ref{constob}, in the absence of a bound on $h_L$ and $h_R$,
    the physical algebras will be of Type II$_\infty$ rather than Type II$_1$.

The final step in deriving those algebras is to impose the isometry group of the full Schwarzschild spacetime as a set of gauge  constraints. The relevant isometry group consists of a single copy of $\R_t$. Because this isometry acts nontrivially at the asymptotic boundary, the correct constraint is not that the bulk isometry generator $\hat h$ annihilates all physical states, but rather that the generator is equal to the corresponding generator in the asymptotic isometry group. In this case, the  constraint is exactly \eqref{ADMconstraint}, namely
\be
\hat h = h_R - h_L
\ee
This is the same constraint that we found for observers in de Sitter, except that the energy of the observers has been replaced by \emph{minus} the subtracted ADM energies. 

The problem of how to incorporate such a constraint on the Hilbert space was already discussed in section \ref{second}. It leads to the elimination of one of the two factors of $L^2(\R_t)$.

The physical algebra of operators in the right exterior consists of the invariant subalgebra of the extended algebra of operators $\A_r^{\mathrm{ext}}$ given in eqn. \eqref{eq:extendedalgebra}. This algebra is generated by $h_R$ together with $ e^{\i t_R \h h}\A_{r,0} e^{-\i t_R \h h}$.
Operators $ e^{\i t_R \h h}\A_{r,0} e^{-\i t_R \h h}$ are
 bulk operators that have been dressed to the right boundary. Similarly the algebra $\A_r^{\mathrm{ext}}$ is generated by $h_L$ together with $ e^{-\i t_L \h h}\A_{l,0} e^{\i t_L \h h}$. In the extended Hilbert space description, no operators that act solely on $\mathcal{H}_0$ are gauge-invariant; any gauge-invariant operator must be dressed to either the right or the left boundary (or to some combination of the two).

\subsection{Boundary Interpretation}

If the Schwarzschild spacetime $Y$ is replaced by AdS-Schwarzschild, then the discussion above is completely unchanged.  However in this case we know that the bulk gravity theory has a holographic dual. This holographic dual has an associated parameter $N$, such that the $G_N \to 0$ limit corresponds to $N \to \infty$. At finite $G_N$, the left and right boundary algebras are Type I. We will see that the Type II$_\infty$ algebras describe a particular large $N$ limit of these Type I algebras.

Let us first understand the boundary interpretation of the Hilbert space $\mathcal{H}$ described in \eqref{eq:Hphys}. For the purposes of illustration, it is helpful to consider the state
$$
\ket{\tilde \Psi} = \int \d t f(t) \ket{\Psi_{HH}(t)}~,
$$
where $\ket{\Psi_{HH}(t)} = \exp(\i h_L t) \ket{\Psi_{HH}}$ is a timeshifted version of the Hartle-Hawking state. In other words, $\ket{\tilde \Psi}$ is the product of $\ket{\Psi_{HH}} \in \mathcal{H}_0$ with some function $f(t) \in L^2(\R_t)$ that describes a state of the  the time-shift mode. The holographic dictionary tells us that the boundary dual of $h_L$ is $H_L - M_0$, where $H_L$ is the left boundary Hamiltonian, and the dual of $\ket{\Psi_{HH}}$ is the thermofield double state
\be
\ket{\Psi_\TFD} = \sum_i e^{-\beta E_i/2} \ket{i}_L \ket{i}_R~,
\ee
where the sum is over all energy eigenstates $\ket{i}$. Hence
\be \label{eq:microTFD}
\ket{\tilde \Psi} = \sum_i  \int \d t f(t)  e^{\i (E_i - M_0) t -\beta E_i/2} \ket{i}_L \ket{i}_R = \sum_i \tilde f(E_i - M_0) e^{-\beta E_i/2} \ket{i}_L \ket{i}_R.
\ee
In other words, $\ket{\tilde \Psi}$ is dual to a variant of the thermofield state, with finite energy fluctuations that are controlled by the Fourier transform $\tilde f(E_i - M_0)$ of $f(t)$.

Note that to construct the thermofield double state itself we would need to have $f(t) = \delta(t)$. Since $\delta(t) \not\in L^2(\R_t)$, the large $N$ limit of $\ket{ \Psi_\TFD}$ does not lie in $\mathcal{H}$. Indeed, in the strict $N \to \infty$ limit, the left and right boundary algebras of any Hilbert space that contains the thermofield double cannot be Type I or II, because the fluctuations in the modular Hamiltonian of the state $\ket{\Psi_\TFD}$ diverge at large $N$. If one instead works in a perturbative expansion in $1/N$, then the thermofield double state  can be described by a Gaussian wavefunction for the timeshift mode where the energy fluctuations are of order $N$  and hence $\Delta t = O(1/N)$ \cite{GCP}. 

What about states in $\mathcal{H}$ where the QFT degrees of freedom are not in the Hartle-Hawking state? The holographic dictionary tells us that QFT operators in the large $N$ limit are dual to single-trace boundary operators. In fact, the full Hilbert space $\mathcal{H}$ is spanned by states that can be constructed using finite products of single-trace operators from states of the form \eqref{eq:microTFD}. This is discussed in detail in \cite{CPW}. The algebras associated to the left and right exteriors are generated by the subtracted Hamiltonians $H_{L/R} - M_0$ and single-trace boundary operators.

From a boundary perspective, it is not so surprising that these algebras are Type II, and hence have a finite renormalized trace. At finite $N$, operators whose matrix elements decay sufficiently rapidly at large energy will have a finite Hilbert space trace. For simple operators that are tightly peaked in energy in a finite window around $M_0$,  this trace should diverge at large $N$ as $\exp(S_{BH})$ times some $N$-independent constant. The Type II algebra trace simply divides out the divergent constant $\exp(S_{BH})$, and keeps the finite piece.

\vskip1cm

\noindent{\it Acknowledgments} VC
is supported by a grant from the Simons Foundation (816048, VC).
  GP is supported by the UC Berkeley physics department, the Simons Foundation through the ``It from Qubit" program, the Department of 
  Energy via the GeoFlow consortium (QuantISED Award DE-SC0019380) and AFOSR (FA9550-22-1-0098); he also acknowledges support from an 
  IBM Einstein Fellowship at the Institute for Advanced Study.
RL acknowledges partial support by MIUR FARE R16X5RB55W
QUEST-NET, GNAMPA-INdAM and the MIUR Excellence Department Project awarded to
the Department of Mathematics, University of Rome Tor Vergata, CUP E83C18000100006.
 Research of EW is partly supported by NSF Grant PHY1911298.
 We thank D. Marolf for
remarks on quantization in de Sitter space.

  \appendix

\section{Crossed Product and Takesaki Duality}\label{takesaki}

For the convenience of the reader, we recall here a few basic facts on the structure of Connes and Takesaki's crossed product of a von Neumann algebra by a group action. The crossed product construction is analogous to the semi-direct product in group theory. 
 We refer to \cite{T, vD} for a complete exposition.  
 \smallskip

\noindent
{\it Crossed product.} 
Let $M$ be a von Neumann algebra on a Hilbert space $\H$, $G$ a locally compact group, and $\al$ an action of $G$ on $M$; by changing the representation, we may assume that $\al$ is implemented by a unitary action $U$ of $G$ on $\H$, namely $\al =\Ad\, U$. The {\it crossed product} $\hat M = M\rtimes_\al G$ of $M$ by $\al$ is the von Neumann algebra on $\H\otimes L^2(G)$ 
\be
\hat M = \{x\otimes 1, u(g) : x\in M, g\in G\}''
\ee
(double commutant), where
$u(g) = U(g)\otimes\l(g)$, with $\l$ the left translation unitary representation of $G$ on $L^2(G)$. Namely, $\hat M$ is the von Neumann algebra generated by $M\otimes 1$ and $u(g), \, g\in G$. Note that $\Ad \,u = \al$ on $M$. The construction of $\hat M$ does not depend on the choice of the unitary implementation $U$ of $\al$, up to a natural isomorphism. 

Let $\beta: G\to {\rm Aut}\,M$ be another action. We shall say that $\al$ and $\beta$ are {\it cocycle equivalent} if there exists a unitary $\al$-cocycle $g\in G \mapsto w(g)\in M$, namely $w(gh) = w(g)\al_g(w(h))$, $w(\cdot)$ continuous map with $w(g)$ unitary, such that $\beta_g(x) = w(g)\al_g(x)w(g)^*$.  If $\al$ and $\beta$ are cocycle equivalent, there exists a natural isomorphism
\be
\{M\rtimes_\al G, \hat \al\} \simeq \{M\rtimes_\beta G, \hat \beta\}\, ;
\ee
this isomorphism is the identity on $M$ and maps $u(g)$ to $w(g)u(g)$. 
\smallskip

\noindent
{\it Dual system.} 
In the following, we assume that $G$ is abelian, denote by $\hat G$ the dual group and by $\langle p,g\rangle$, $p\in \hat G, g\in G$, the duality pairing between $\hat G$ and $G$. In the particular, if $G = \mathbb R$, we have $\hat G = \mathbb R$ and $\langle s,t\rangle = e^{\i ts}$. 

Let $V$ be the unitary representation of $\hat G$ on $L^2(G)$, with $V(p)$ the multiplication by $\langle p,\cdot\rangle$ on $L^2$-functions. Then $v = 1\otimes V$ implements
the action $\hat \al$ of $\hat G$ on $\hat M$ given by
\be
 \hat\al_p(x) = x,\quad \hat\al_p(u(g)) = \langle p,g\rangle u(g),\quad  g\in G,\, p\in \hat G,\, x\in M,
\ee
(with the identification $x = x\otimes 1$) called  the {\it dual action}. The $\hat\al$-fixed point subalgebra $M^{\hat\al}$ is $M$:
\be
M = \{x\in\hat M : \hat\al_p(x) = x, p\in \hat G\}.
\ee
Let $\hat{\hat M} = \hat M \rtimes_{\hat \al} \hat G$ be the crossed product of $\hat M$ by $\hat \al$. Then $\hat M$ is the $\hat{\hat \al}$-fixed point algebra
$\hat{\hat M}^{\hat{\hat \al}}$ of 
$\hat{\hat M}$:
\be
\hat M = \{x\in \hat{\hat M} : \hat{\hat \al}(x) = x\}.
\ee
\noindent
{\it Takesaki duality.} 
Takesaki duality shows the isomorphism of the dynamical systems
\be
\{\hat{\hat M},  \hat{\hat \al}\} \simeq \{M\otimes B(L^2(G)),\al\otimes \Ad \,\l\}
\ee
($\Ad\, \l$ is the action on $ B(L^2(G))$ implemented by $\l$),
namely $M\otimes  B(L^2(G))$ is isomorphic to $\hat{\hat M}$ with an isomorphism
that interchanges $\alpha\otimes \Ad\lambda$ with
 with $\hat{\hat \al}$. Therefore
\be
\big(M\otimes  B(L^2(G))\big)^{\al\otimes {\rm Ad}\,\l} \simeq M\rtimes_\al G \, .
\ee
\smallskip

\noindent
{\it Dominant and integrable actions.} 
Recall that $M$ is properly infinite if $M$ contains a type I$_\infty$ factor (for example, if $M$ is of type III); equivalently, if $M$ is isomorphic to $M\otimes F$ with $F$ a type I$_\infty$ factor. 

By definition, the action $\al$ on $M$ is {\it dominant} if $M^\al$ is properly infinite and
\be
\{M,\al\} \simeq \{M\otimes  B(L^2(G)), \al\otimes \Ad\,\l\}
\ee
(isomorphism of dynamical systems). 
Then
\be
\al\ {\rm dominant} \implies M^\al \ \text{is isomorphic to} \ \hat M\, .
\ee
$\al$ is dominant iff there exist unitaries $z(p)\in M$, $p\in \hat G$ (Borel family)
such that 
\be
\al_g(z(p)) = \langle p, g\rangle z(p), \quad g\in G,\,  p\in \hat G. 
\ee
The dual action is always dominant. 
Every action $\al$  is cocycle equivalent to a dominant action $\beta$. 

The action $\al$ is said to be {\it integrable} if the linear span of
\be
\big\{x\in M^+, \int_G \al(x)\,{\mathrm d}g < \infty \big\} 
\ee
is weakly dense in $M$. Every dominant action is integrable. If $G$ is compact, every action is integrable. 
Moreover:
\be
\al \ {\rm integrable}\implies M^\al \ \text{is isomorphic to a reduced von Neumann subalgebra of $\hat M$}\, ,
\ee
namely $M^\al \simeq e\hat M e$ for some projection $e$ of $\hat M$. 
\smallskip

\noindent
{\it Weights, operator-valued weights, dual weights.} 
A normal, faithful, semifinite  (n.f.s.) weight $\f$ on $M$ is a map $M_+\to [0,+\infty]$ (the extended positive real numbers), with $M_+$ the cone of positive elements of $M$, such that
\begin{equation}\label{wei}
\f(x+y) = \f(x) + \f(y),\quad \f(ax) = a\f(x),\quad x,y\in 
M_+,\, a\geq 0\,,
\end{equation}
with $\sup \f(x_i) = \f(\sup x_i)$ for every bounded, increasing nets $\{x_i\}$ in $M_+$ 
(normality), $\f(x) = 0 \implies x=0$, $x\in M_+$ (faithfulness), and
the linear span $M_0$ of the $x\in M_+$ such that $\f(x) < \infty$ is dense in $M$ (semifiniteness). $\f$ can be extended by linearity to all $M_0$. 

If $N\subset M$ is a von Neumann subalgebra, a n.s.f. {\it operator valued weight} $E: M_+ \to \tilde N_+$ is defined similarly as in \eqref{wei}, with $\tilde N$ the extended positive part of $N_+$ 
(see \cite{T} for the definition of $\tilde N_+$), with the property
\be
E(n^*xn) = n^* E(x) n\, ,\quad x\in M_+,\, n\in N\, .
\ee
The notion of operator-valued weight, due to Haagerup, generalises the notion of conditional expectation.  

Let $\hat M = M\rtimes_\al G$ be as above. Then
\be
T(x) = \int_{\hat G} \hat\al_p(x)\,{\mathrm d}p\, ,\quad x\in \hat M_+\, ,
\ee
defines a n.f.s. operator-valued weight $T: \hat M_+ \to \tilde M_+$.

Every n.f.s. weight $\f$ of $M$, in particular every normal, faithful state $\f$ on $M$, has a canonical lift to a n.f.s. weight $\hat \f$ on $\hat M$
given by
\be
\hat \f = \f\cdot T \, ;
\ee
$\hat \f$ is called the {\it dual weight} of $\f$. 

Note that, if $G$ is not discrete, $\hat\f(u(t))$ is not defined; however, formally, $\hat\f(u(t))= 0$, $t\neq 0$. 
For example, if $G = \mathbb Z$, then an element $X$ of $\hat M$ has a Fourier series expansion (in some topology)
with Fourier coefficients $x_n\in M$, so
\be
X = \sum_n x_n u(n)\ {\rm and} \ \hat\f(X) = x_0. 
\ee
If $G = \mathbb R$, this is formally true: if $ X = \int x(t)u(t)\,{\mathrm d}t$, with $x(t)$ a ``good" $M$-valued function, then $\hat\f(X) = \f(x(0))$. 

Note also that, if $M = \C$, then $\hat M$ is $L^\infty(\mathbb R)$ and $\hat \f$  is the mean in Fourier transform, namely $\hat\f(\lambda(f)) = f(0)$, say for a smooth, compactly supported $f$, where $\l(f) = \int f(s)\l(s)\d s$. 

\smallskip

\noindent
{\it Crossed product by the modular group.} We now specialise the above discussion to the case $G = \mathbb R$ and the action $\al$ is given by the modular group $\sigma^\f$ of a faithful normal state $\f$ on $M$  (the discussion may be generalized to the case of n.f.s. weights). Thus
\be
\hat M = M\rtimes_{\sigma^\f} \mathbb R\, .
\ee
By the Connes cocycle Radon-Nikodym theorem, the modular groups associated with different states are cocycle equivalent, therefore $\hat M$ does not depend on the choice of the state $\f$, up to a natural isomorphism. 

We may define the action $\gamma$ of $\mathbb R$ on $\hat M$ by setting
\be
\gamma_s(x) = \sigma^\f_s(x),\ x\in M,\quad \gamma_s(u(t)) = u(t),\quad s,t\in \mathbb R,
\ee
indeed $\gamma_s =\Ad \,u(s)$ on $\hat M$. It turns out that $\gamma$ is the modular group $\sigma^{\hat \f}$ of the dual weight $\hat \f$ of $\f$ on $\hat M$. So $\s^{\hat\f}$ is inner. Thus, there exists a trace $\tau$ on $\hat M$ (a n.f.s. weight $\tau$ such that $\tau(z^* x z) = \tau(x)$ for all unitaries $z\in \hat M$ and $x\in \hat M_+$) given by
\be
\tau(\cdot) = \hat\f(\rho^{-1}\cdot) \, ,
\ee
where $u(t) = \rho^{\i t}$;  $\rho$ is affiliated to $\hat M$, the density matrix of $\hat \f$ with respect to $\tau$. 
Therefore, $\hat M$ is {\it semifinite}. 

With $\vartheta = \hat \sigma^\varphi$ the dual action of $\sigma^\varphi$ on $\hat M$, we have $\vartheta_s (u(t))= e^{\i ts} u(t)$, 
so $\vartheta_s (\rho) = e^{s} \rho$, therefore
\be
\tau\cdot\vartheta_s = e^{-s} \tau\, ,
\ee
namely $\vartheta$ is a trace scaling action on $\hat M$. 

If $M$ is a type III von Neumann algebra, then $\hat M$ is a type II$_\infty$ von Neumann algebra. If $M$ is a type III$_1$ factor, then $\hat M$ is a II$_\infty$ factor. 

Let $x\in M$ and $f$ a Borel function on $\mathbb R$ such that $\int e^{-s}| f(s)|{\mathrm d}s < \infty$. Since $\rho^{\i t} = U(t)\otimes \lambda(t)$, we have
\be
\tau(x f(\log\rho)) = \hat \f(x \rho^{-1} f(\log\rho))  
=   \f(T(x \rho^{-1} f(\log\rho))) \ee
\be\notag
~~~~~~~~~~~~~~~~~~~=   \f(x)T( \rho^{-1} f(\log\rho))= \f(x)\int e^{-s}  f(s){\mathrm d}s.
\ee
In particular, if $ f$ is the characteristic function of $[0,\infty)$, the projection $e = f(\log\rho)$ has finite trace: $\tau(e) = \int_0^\infty e^{-s}{\mathrm d}s = 1 < \infty$.

  \section{The Hilbert Space of Coinvariants}\label{coinvariants}
  
  In this appendix, we will review the construction \cite{higuchi,marolf} of a physical Hilbert space for quantum fields 
  coupled to gravity in de Sitter space, in the limit $G\to 0$.  
  Then we will explain how the construction can be naturally placed in the language of BRST cohomology.  
  
  \subsection{Constraints and Group Averaging}\label{averaging}
  
  First, we recall the construction of a Hilbert space for a quantum field in de Sitter space $X$, in the absence of gravity \cite{CT,SS,BD,Mo,Al}.  For illustration,
  we consider the case of a free scalar field $\phi$ of mass $m$, with action
  \be\label{scalaction}I=-\frac{1}{2}\int\d^D x \sqrt{\det g} \left(g^{ij}\partial_i\phi\partial_j\phi +m^2\phi^2\right).\ee
  $\phi$ obeys the massive Klein-Gordon equation
  \be\label{kgeqn}\left(-g^{ij}D_i D_j +m^2\right)\phi=0. \ee
By $\Psi_\dS$, we mean a state with Gaussian correlations  in which the one-point function $\la\phi(x)\ra$ vanishes, and
the two-point function $\la\phi(x)\phi(y)\ra$ is determined by analytic continuation from Euclidean signature.   A Hilbert space $\H$ that provides an
irreducible representation of the canonical commutation relations of the field $\phi$ and contains such a state $\Psi_\dS$ can be constructed in an essentially
unique way.   For this, we observe first that $\H$ must contain states obtained by acting on $\Psi_\dS$ with any number of $\phi$ fields.   So $\H$ must contain states
\be\label{zeqn}\Psi_f = \int \d x_1\d x_2 \cdots \d x_n f(x_1,x_2,\cdots , x_n) \phi(x_1) \phi(x_2)\cdots \phi(x_n) |\Psi_\dS\ra\ee
where $f(x_1,x_2,\cdots, x_n)$ is any compactly supported function of an $n$-plet of  points\footnote{Since Wightman functions are distributions in Lorentz
signature, there is no need to require that the points $x_1,\cdots, x_n$ be distinct.}  $x_1,x_2,\cdots,$ $x_n\in X$.  
Let $\H_{0,n}$ be the space generated by all such states $\Psi_f$, where $f$ depends on $n$ points, and let $\H_0=\oplus_{n=0}^\infty \H_{0,n}$.    The vector $\Psi_\dS$ that we started with is an element of 
$\H_{0,0}\subset \H_0$.  
Having specified the correlation
functions of the field $\phi$ in the state $\Psi_\dS$, we immediately know what should be the inner products of states  in $\H_0$.
For $\Psi_{f_1}\in \H_{0,n}$, $\Psi_{f_2}\in \H_{0,m}$, we define
 \be       \label{weqn}\la \Psi_{f_1},\Psi_{f_2}\ra=\int \d x_1\cdots \d x'_m \bar f_1(x_1,\cdots, x_n) f_2(x'_1,\cdots, x'_m)
\la\Psi_\dS|\phi(x_n)\cdots \phi(x_1)\phi(x'_1)\cdots \phi(x'_m)|\Psi_\dS\ra. \ee
These integrals always converge, because $f_1$ and $f_2$ were defined to have compact support.
With this inner product, in $\H_0$ there are many null vectors; for example, if  $f=(-g^{ij}D_i D_j+m^2)\t f$ for some function $\t f$ of compact support, 
then $\Psi_f$ is a null vector,
because of the Klein-Gordon equation satisfied by $\phi$.
  Taking the quotient of $\H_0$ by its null vectors, we get a space $\H_1$ that
has all the properties of a Hilbert space except completeness.  Its Hilbert space completion is the Hilbert space $\H$ of the massive scalar field in de Sitter space.

Now let us try to couple to gravity, in the limit $G_N\to 0$.     After turning on gravity, we cannot just consider the fluctuations of matter fields;
we have to also construct a Hilbert space of small fluctuations in the metric. This does not change the picture in a qualitative way.
What does qualitatively change the picture is that in the presence of gravity, one wishes to impose the symmetry group $G_\dS$ as a group of constraints.
This is subtle.   Naively, for $G_\dS$ to be a group of constraints means that states should be $G_\dS$-invariant.  So one's first thought is to impose the constraints
by restricting from $\H$ to its $G_\dS$-invariant subspace $\H^{G_\dS}$.    However, 
there are actually no $G_\dS$-invariant vectors in $\H$ except  multiples of $\Psi_\dS$.  
The definition of $\H$ started with a space $\H_0$ constructed using a function $f$ of compact support on the product of $n$ copies of $X$.   The compact support condition trivially
means that such states cannot be $G_\dS$-invariant for $n>0$.   To define $\H$, we then took a completion, so that states in $\H$ are not necessarily derived
from functions of compact support.   Nevertheless, states in $\H$ are never $G_\dS$-invariant.  One can define a vector space of $G_\dS$-invariant
states $\Psi_f$, where $f(x_1,x_2,\cdots, x_n)$ is $G_\dS$-invariant, but there is no way to define an inner product on the space of such states, since
the necessary integral never converges.

 Therefore a different procedure is needed.   The procedure that works is
to consider not invariants in the action of $G_\dS$ on the vectors $\Psi_f$, but coinvariants.    For any $\Psi_f\in\H_0$ and any $g\in G_\dS$, we simply declare
$\Psi_f$ and $g\Psi_f$ to be equivalent.  The equivalence classes are called coinvariants.
  We also impose a constraint that $\la \Psi_\dS|\Psi_f\ra=0$ (the purpose of this constraint is to avoid a divergent disconnected contribution in
eqn. (\ref{yoko}) below).  The space of $\Psi_f$ subject to the equivalence relation and the constraint 
 defines a new space $\H'_0$,  on which $G_\dS$ acts trivially by fiat.
 The formula of eqn. (\ref{weqn}) does not define an inner product on $\H_0'$, because in general for $\Psi_{f_1}, \,\Psi_{f_2}\in\H_0$ and $g\in G_\dS$, 
$\la\Psi_{f_1},\Psi_{f_2}\ra\not=\la\Psi_{f_1},g\Psi_{f_2}\ra$.  However, averaging or more precisely integrating over $G_\dS$ gives
a simple fix.   We define 
\be\label{yoko} \left( \Psi_{f_1},\Psi_{f_2}\right) =\int_{G_\dS} \d g\, \la \Psi_{f_1},g \Psi_{f_2}\ra,\ee
where $\d    g$ is the invariant Haar measure on the $G_\dS$ group manifold.
It is fairly obvious that if the integral converges, the inner product $(~,~)$  is consistent with the equivalence relation (any state of the form $(1-g)\Psi$ is a null vector
for this inner product), and therefore the integral defines an inner product on $\H_0'$.
The convergence of the integral is subtle; see for example \cite{marolf} for analysis.  Also subtle and not fully understood is the question of whether the inner product
defined this way is positive semi-definite. This is known to be the case in some important examples \cite{higuchi}.  
We will not address these issues here; our goal is only to explain how group averaging can be interpreted in terms of
BRST quantization.
Assuming  the inner product $(~,~)$ exists and  is positive semi-definite on the space $\H_0'$, 
we can proceed  as before.  We define a space $\H_1'$ as the quotient of $\H_0'$ by null vectors.
The Hilbert space completion of $\H_1'$ is then the desired Hilbert space $\h\H$ taking into account the gravitational constraints.   

  \subsection{The BRST Interpretation}\label{BRST}

So far, we have just summarized the construction of \cite{higuchi,marolf}, which has been called ``group averaging'' because of the role of integration over $G_\dS$
in defining the inner product.   In general, group averaging can be naturally understood in a framework\footnote{See \cite{Henneaux} for
a relatively short and highly readable introduction to the BRST and BV methods of quantization.}  of BRST and BV quantization
\cite{shvedov}.  Our goal in the rest of this appendix is to explain this point.  BRST quantization (rather than the more
general BV quantization) is adequate when the constraints generate a Lie algebra, so this  approach will suffice.

Let $t_p$, $p=1,\cdots, \dim\,G_\dS$ be the linear operators that generate the action of $G_\dS$ on $\H_0$.  They obey 
\be\label{comrels} [t_p,t_q]=f_{pq}^r t_r, \ee
where $f_{pq}^r$ are the structure constants of $G_\dS$.   Those structure constants obey the Jacobi identity $f_{[pq}^s f^t_{r]s}=0$, or in more detail
\be\label{jacobi} f_{pq}^s f^t_{rs} +  f_{qr}^s f^t_{ps}   +f_{rp}^s f^t_{qs}     =0 .\ee
Introduce fermion operators $c^r$ and $b_s$, known respectively as ghosts and antighosts, with anticommutation relations
\be\label{anticom}\{c^r,b_s\}=\delta^r_s, ~\{c^r,c^s\}=\{b_r,b_s\}=0. \ee
Ghost number is defined so that $c^r$ has ghost number 1 and $b_s$ has ghost number $-1$.   The antighost operators have an irreducible representation on
a finite-dimensional vector space $\K$ that contains a state $|\negthinspace\downarrow\ra$ of minimum ghost number, with $b_s|\negthinspace\downarrow\ra=0$. Other
states in $\K$ are obtained by acting on $|\negthinspace\downarrow\ra$ with a polynomial in the $c$'s.  In particular  $\K$  also contains the state $|\negthinspace\uparrow\ra=c^1 c^2\cdots c^{\dim G_\dS}|\negthinspace\downarrow\ra$ of maximum ghost number, annihilated by the $c$'s.   Mathematically, an additive constant in the ghost number is usually fixed so that 
$|\negthinspace\downarrow\ra$ has ghost number zero and therefore $|\negthinspace\uparrow\ra$ has ghost number $\dim\,G_\dS$.   
We will use that convention, though physically it is not always the most natural choice.

The group $G_\dS$ acts on the combined space $\H_0\otimes \K$ with generators
\be\label{newgen}T_r=t_r- \sum_{s,t} f_{rs}^t c^s b_t. \ee
The BRST operator is defined as 
\be\label{brst}Q=\sum_r c^r t_r -\frac{1}{2}\sum_{r,s,t} f_{rs}^t c^r c^s b_t.\ee
The factor of $1/2$ is important in verifying the crucial relation
\be\label{nilp} Q^2=0,\ee
which enables one to define the cohomology of $Q$.   The cohomology of $Q$ at ghost number $n$ is defined as the space of states $\Psi$ of ghost number $n$
that satisfy $Q\Psi=0$, modulo those of the form $\Psi=Q\chi$. We denote this cohomology group as $H^n(Q,\H_0)$;   mathematically, it  is called the degree $n$ Lie algebra cohomology of $G_\dS$ with
values in $\H_0$. The factor of $1/2$ is also important in verifying  the anticommutation relation  
\be\label{anticomm} \{Q,b_r\}=T_r. \ee
This anticommutation relation ensures that the group generators act trivially on the cohomology of $Q$, since if $Q\Psi=0$,
then $T_r\Psi =\{Q,b_r\}\Psi=Q(b_r\Psi)$.

In the BRST approach to quantization, the space of physical states is defined via the cohomology of the BRST operator $Q$ at a specific value of the ghost number.
Which is the correct value is somewhat theory-dependent.   However, normally there is one specific ``allowed'' value of the ghost number and physical states
are defined in terms\ of the BRST cohomology\footnote{A completion is involved, as discussed momentarily.} at that value of the ghost number.    If one did not
have this restriction to a particular value of the ghost number, there would be an invariant distinction between physical states based on their ghost number.

In the example of the de Sitter space with weakly coupled gravity, there are two important values of the ghost number, namely the minimum and maximum possible
values.   A state in $\H_0\otimes \K$ of the minimum possible ghost number is of the general form $\h\Psi=\Psi \otimes |\negthinspace\downarrow\ra$, with $\Psi\in\H_0$.
The condition $Q\h\Psi=0$ reduces to $t_a\Psi=0$, in other words $\Psi$ must be $G_\dS$-invariant.   This is the naive approach to imposing the constraints.
As explained earlier, it is unsatisfactory, because except for $\Psi_\dS$ itself, there are no $G_\dS$-invariant states in $\H_0$ or in its Hilbert space completion.
On the other hand, consider states with the maximum possible ghost number.   Such a state has the form $\h\Psi=\Psi\otimes |\negthinspace\uparrow\ra$.
For such a state, the condition $Q\h\Psi =0$ is trivial, since the ghost number of $Q\h\Psi$ exceeds the maximum possible value.   However,
we need to consider the equivalence relation $\h\Psi\cong \h\Psi+Q\chi$.    Concretely, a general state of the appropriate ghost number is
 $\chi=\sum_a \chi^a b_a|\uparrow\ra$ for some states
$\chi^a\in\H_0$. For this choice, the equivalence relation becomes\footnote{This formula and some previous ones need slight modification if $G$ is not unimodular.  
See footnote \ref{unimodular} and \cite{MarolfReview}.  The symmetry groups of de Sitter space and of the static patch are unimodular.}
\be\label{eqrel} \Psi\cong \Psi+\sum_a t_a\chi^a. \ee
Thus, for states of maximum ghost number, the equivalence relation reduces to an equivalence relation on $\H_0$, namely
 $t_a\chi\cong 0$ for any state $\chi\in\H_0$ and for $a=1,\cdots,\dim\,G_\dS$.  This
 has the same content as the equivalence relation $(1-g)\chi\cong 0$ that we used in discussing group averaging, since
from $(1-g)\chi\cong 0$, by differentiating with respect to $g$ at $g=1$ we can deduce that $t_a\chi\cong 0$;
conversely, from $t_a\chi\cong 0$ for all $\chi$, we can deduce $(1-g)\chi\cong 0$ by integration.   
So in short the top degree cohomology of the BRST operator $Q$ acting on $\H_0\otimes \K$ is the starting point in the construction of the Hilbert space via group
averaging.  

The rest of the discussion proceeds  rather as before, defining an inner product, discarding null vectors, and taking a completion to get a Hilbert space.   In the BRST
language, the construction of the inner product is naturally described as follows.

First
 define an indefinite  hermitian inner product on $\K$ such that $\la\downarrow|\uparrow\ra=1$; $c^r$, $b_s$ are hermitian;
and $\la\Psi,\Psi'\ra=0$ if $\Psi,\Psi'$ have ghost number $n,n'$ with $n+n'\not= \dim\,G_\dS$.
These conditions completely determine the inner product; for instance $\la c^r \negthinspace\downarrow| b_s\negthinspace\uparrow\ra=\la\downarrow|c^r b_s|\downarrow\ra=\delta^r_s$, since
$\{c^r,b_s\}=\delta^r_s$ and $c^r|\negthinspace\uparrow\ra=0$.     
On $\H_0'\otimes \K$, define an inner product  $\la~,~\ra$ that is the  tensor product of the inner product just defined on $\K$ and the natural one on $\H_0'$ that was defined in eqn. (\ref{weqn}).
Then $Q$ is hermitian in this inner product, so  $\la\Psi,\Psi'\ra$, for $\Psi$, $\Psi'$ of respective ghost numbers $n$ and $\dim\,G_\dS-n$, induces a natural
sesquilinear  map 
\be\label{zimbox} \la~,~\ra:H^n(Q,\H_0')\times H^{\dim\,G_\dS-n}(Q,\H_0')\to \C.\ee    However, this map is not very useful for our purposes, since physical states
lie in $H^{\dim\,G_\dS}(Q,\H_0')$, which is paired with $H^0(Q,\H_0')=0$.  

There is a simple way around this.   We defined $\H_0$ in terms of states $\Psi_f$ where $f$ has compact support.  We can drop the compact support
condition and define a ``dual'' space $\t\H_0$ generated by states $\Psi_f$ with any smooth function $f$, not necessarily of compact support.
The inner product $\la~,~\ra$ defined in eqn. (\ref{weqn})  by integration over a product of copies of $X$ makes sense as a pairing between $\t\H_0$ and $\H_0$.
(It does not make sense as a pairing between $\t\H_0$ and itself, which is why it is necessary to base the construction on the cohomology with values in $\H_0$.)  
We can define the BRST cohomology $H^n(Q,\t\H_0)$ for states valued in $\t\H_0$, and in particular $H^0(Q,\t\H_0)$ is the naive space of  unnormalizable
states $\Psi_f$ with $G_\dS$-invariant
$f$.  As we explained at the beginning, the only trouble with that space is that it is not a Hilbert space in a natural way.  

However, the pairing between $\t \H_0$ and $\H_0$ that comes from eqn. (\ref{weqn}) is useful.  Taking the tensor product of this
with the inner product on $\K$, we 
get a natural sesquilinear pairing $\la~,~\ra: (\t \H_0\otimes \K) \times (\H_0\otimes \K)\to \C$ and, since $Q$ is again hermitian, this 
leads to a much more interesting map 
\be\label{oregon} \la~,~\ra: H^n(Q, \t \H_0)\times H^{\dim\, G_\dS-n}(Q,\H_0') \to \C.\ee
Therefore, to define a hermitian inner product on $H^{\dim\,G_\dS}(Q,\H_0')$, we  need a linear map $\eta:H^{\dim \,G_\dS}(Q,\H_0')\to  H^0(Q,\t\H_0') $ (sometimes called a rigging map in the literature).
Such a map is the composition of $b_1 b_2\cdots b_n$ with integration over the action of $G_\dS$.  The integration operation can be formally denoted
$\delta(\vec T)$, since it projects onto (unnormalizable) states that are annihilated by the $T$'s.   Thus $\eta=b_1 b_2 \cdots b_{\dim\,G_\dS}\delta(\vec T)$.
We have\footnote{The expression 
$[Q,\eta\}$, for any operator $\eta$,  is defined as $Q\eta-(-1)^\eta\eta Q$, where $(-1)^\eta$ is 1 or $-1$ if $\eta$ is bosonic or fermionic.  In the present case,
$\eta$ is bosonic or fermionic depending on whether $\dim\,G_\dS$ is even or odd.} 
 $[Q,\eta\}=0$, since\footnote{\label{unimodular}The formula $\{Q,b_r\}=T_r$ implies that $\[Q,b_1 b_2\cdots b_n\}$ is a linear combination of operators
 $b_1 \cdots b_{i-1} T_i b_{i+1}\cdots b_n$.  In such an expression, we move $T_i$ to the right, where it annihilates $\delta(\vec T)$.   In moving
 $T_i$ to the right, we pick up terms that are of order $n-1$ in $b$, since 
$[T_i,b_j]=f^k_{ij} b_k$.  Summing all such contributions, we get a multiple of
$\epsilon^{i_1 i_2\cdots i_n}f_{i_1 i_2}^k b_k b_{i_3}\cdots b_{i_n}$ where $\epsilon^{i_1 i_2\cdots i_n}$ is completely antisymmetric.
This expression vanishes for  $G_\dS$ and also for the symmetry group of the static patch.   In fact, $\epsilon^{i_1 i_2\cdots i_n}f_{i_1 i_2}^k b_k b_{i_3}\cdots b_{i_n}$ 
is a multiple of $ f^t_{rt}\{c^r, b_1 \dots b_n\}$, and vanishes if and only if $f^t_{rt}=0$, which is the condition for $G$ to be unimodular (meaning that a left-invariant measure
on the group manifold is also right-invariant).
If $G$ is not unimodular, the group averaging procedure must be slightly modified by correcting the group generators \cite{MarolfReview}.  Any semisimple Lie group and
any group that has no invariant subgroup of codimension 1 is unimodular. (The symmetry group of the static patch is unimodular though it has such a subgroup.)  A simple example
of a Lie group that is not unimodular is the group of $2\times 2$ upper triangular matrices.  
}
$\{Q,b_r\}=T_r$, $[Q,T_r]=0$, and $T_r\delta(\vec T)=0$.  So $\eta$ does give a natural map from $H^{\dim\,G_\dS}(Q,\H_0')$ to
$H^0(Q,\t\H_0')$.   In short, then, from a BRST point of view, the natural hermitian inner product on $H^{\dim\,G_\dS}(Q,\H_0')$ is
\be\label{zoltan} (\Psi,\Psi')=\la\Psi|\eta|\Psi'\ra. \ee
Once this inner product is defined, as usual one reduces by null vectors and takes a completion to get the desired Hilbert space $\h\H$.

One further subtlety  is the role of the original de Sitter state $\Psi_\dS$.  When we summarized the group averaging procedure in Appendix \ref{averaging},
in addition to imposing an equivalence relation that had the same content as $t_a\chi\cong 0$, we also split off $\Psi_\dS$ by restricting to states $\Psi_f$ that
are orthogonal to $\Psi_\dS$.  This was necessary for convergence of the integral (\ref{yoko}).   Similarly, in the BRST approach, $\eta$ only makes sense
as an operator on $\H_0'$, not on $\H_0$.   However, physically it makes sense that $\Psi_\dS$ should be included as a vector in the Hilbert space
$\h\H$ after imposing constraints.  
We believe that a heuristic explanation is as follows.   First of all,
the state $\Psi_\dS |\negthinspace\uparrow\ra$ is contained in the BRST cohomology with values in $\H_0\otimes \K$, so the question is how to define its norm.
Before imposing the $G_\dS$ constraints, the state $\Psi_\dS$ is normalizable and one can normalize it to have norm 1.
 When one imposes the $G_\dS$ constraints, one should divide the norm of the $G_\dS$-invariant state $\Psi_\dS$ by $\vol(G_\dS)$, the (infinite) volume of
 the de Sitter group. However, when one integrates over the group to define the inner product, in the case of a $G_\dS$-invariant state, this just gives back a
 factor of $\vol(G_\dS)$.   In other words, precisely because the Hilbert space inner product is defined by integration over the group, the state $\Psi_\dS$
 has a nonzero norm after imposing the constraints and survives as a vector in the physical Hilbert space.   Obviously, it would be desirable to have an explanation of this
 point that does not involve canceling the infinite factors of $\vol(G_\dS)$.   

\subsection{The Observables}\label{observables}

In BRST quantization, though the ghosts are important in understanding the physical states, the analysis of the observables
is more straightforward.  This has been implicitly assumed in the present article, as our main focus has been the algebra of observables in the static patch, and
we analyzed this algebra with no mention of the ghosts.  The fact that the description of the observables is straightforward can be explained as follows.
 First of all, in the BRST approach, an observable is defined to be an operator $\O$ on the full state space including the ghosts
  that commutes with $Q$ in the $\Z_2$-graded sense: $[Q,\O\}=0$.   This condition ensures that if $Q\Psi=0$,
then $Q\O\Psi=0$, and also that  if $\Psi=Q\chi$, then $\O\Psi=\O Q\chi =(-1)^{\O} Q(\O\chi)$.   The two statements together imply that $\O$ has a well-defined
action on the cohomology of $Q$.   Moreover, if $\O=[Q,\U\}$ for some operator $\U$, then $\O$ acts trivially on cohomology, since if $Q\Psi=0$ then $\O\Psi=
[Q,\U\}\Psi=Q(\U\Psi)$.   So observables in BRST quantization
are defined to be cohomology classes of operators; in other words, an observable is an operator $\O$ that satisfies $[Q,\O\}=0$, modulo operators of the form
$[Q,\U\}$.   

 In general, one shows that at least in the context of perturbation theory (which here means perturbation theory in $G_N$) every cohomology class has a representative that can be constructed with no explicit use of $b$. In the present
context, this  means this means that every operator has
a representative constructed only from matter fields and gravity together with the ghost field $c$.   
Usually, one is primarily interested in operators of ghost number zero.   The reason is that as explained in Appendix \ref{BRST}, normally physical states are defined at just one
value of the ghost number (the maximum value, in the case of de Sitter space) and hence an operator of nonzero ghost number has vanishing matrix elements
between physical states.  An operator of ghost number zero, in its representative that does not contain factors of $b$, automatically also 
does not depend on $c$; in other words, such an operator is constructed from matter fields and gravity only.
    If $\O$ is an operator that is constructed without any use of either $b$ or $c$,
then the condition $[Q,\O\}=0$ reduces to $[t_a,\O]=0$.  In  other words, $\O$ must be $G_\dS$-invariant.   

So while it is subtle to impose the de Sitter constraints on physical states, it is straightforward to impose the constraints on operators: an operator that acts
on the constrained system is simply an operator on the unconstrained system that commutes with the constraints.   This was assumed in the main body of the paper. 

If $\Psi_1$, $\Psi_2$ are physical states of the constrained system and $\O$ is an operator of the constrained system, then
\be\label{toffo}\la\Psi_1| g\O g^{-1}|\Psi_2\ra=\la\Psi_1|\O|\Psi_2\ra, \ee
since physical states were defined to satisfy $g^{-1}\Psi_1=\Psi_1,$ $g^{-1}\Psi_2=\Psi_2$ (mod $\{Q,\cdots\}$).   

We have described this construction for the case of imposing constraints under the full group $G_\dS$, but the same ideas apply for imposing constraints
under a subgroup such as the symmetry group $G_P$ of a static patch $P$. That case is actually much more simple.  One has $G_P=\R_t\times \SO(D-1)$, where
$\R_t$ is the group of time translations of the static patch and $\SO(D-1)$ is a group of rotations.
The group $\SO(D-1)$ is compact, which means that there is no analytical difficulty in defining and understanding the BRST cohomology for $\SO(D-1)$ at
either the top or bottom value of the ghost number (and these two cohomology groups are actually naturally isomorphic).   
The group $\R_t$ of time translations of the static patch is abelian.  The group averaging procedure for this abelian group
is straightforward and was actually implemented in section \ref{hilbertspaces}.

A subtlety in all this accounts for a difficulty that we had in section \ref{emptypatch}.   As we have explained, it is possible to
identify operators that commute with a set of constraints without first defining a Hilbert space that these operators act on.
What one gets this way is an algebra $\B_0$, but not a von Neumann algebra.   To define a von Neumann algebra, one has to construct a Hilbert space $\h\H$ that $\B_0$
acts on, and then one can complete $\B_0$ to a von Neumann algebra $\B$ of operators on $\h\H$.   In general,  $\B$  depends on $\h\H$.
In the main example of the present paper, we start with a Hilbert space $\H_\dS\otimes L^2(\R)\otimes K$, where $\H_\dS$ is a Hilbert space of fields in de Sitter space,
$L^2(\R)$ is the Hilbert space of an observer in $P$, and $K$ is the Hilbert space of a possible observer in the complementary patch $P'$.   
An algebra $\A$ of the static patch acts on $\H$.
 We consider a constraint $\h H=H+q-k $, where $H$ generates the time translation symmetry of the patch $P$, $q$ is the Hamiltonian of the observer in $P$, and $k$ acts on $K$.  
There are two cases of interest:
either $K$ is a second copy $L^2(\R)'$ of $L^2(\R)$ and $k=q'$ is the Hamiltonian of an observer in $P'$, or $K=\C$, $k=0$.
 Operators on $\H_\dS\otimes L^2(\R)$ 
that commute with the constraint are $e^{\i p H}\a e^{-\i p H}$ and (bounded functions of) 
$q$, where $p=-\i \d/\d q$ and $\a\in\A$.   These generate an algebra $\B_0$, which does not depend on whether there is an observer present in $P'$.
However, the completion of $\B_0$ to a von Neumann algebra $\B$ does depend on the presence of an observer in $P'$.   In the case $K=L^2(\R)'$, $k=q'$,
$\B$ is the projected
crossed product algebra of Type II$_1$ that we have studied in this article.    In this case, the Hilbert space $\h\H$ of coinvariants, as a representation of $\B$,
has a ``continuous dimension'' in the sense of Murray and von Neumann that can be any positive real number $d$, depending on what we assume for the range of values of $q'$.   But in case $K=\C$, $k=0$, in the limit $G_N\to 0$, by conjugating by $e^{-\i p H}$, we can reduce to the case that the constraint is just $q$ and the Hilbert space
$\h\H$ obtained after imposing the constraint is the original Hilbert space $\H_\dS$.   In that description, $\B_0$ is generated by $\A$ and $H$, and its von Neumann
algebra completion is the Type I algebra of all bounded operators on $\H$.   This is an unreasonable result, since  the algebra of 
operators accessible to an observer in $P$ should not depend
in this way on whether there is an observer in $P'$. Having no observer in the second patch is similar to taking the limit $d\to 0$ for the continuous dimension
of the Hilbert space, but that limit is not possible for an algebra of Type II$_1$; that is why we seem to get a different algebra in the patch $P$ when there is no
observer in $P'$.   We suggested in section \ref{emptypatch} that actually it is not possible in this problem to take a strict limit $G_N\to 0$, and
that  in the absence of an observer in $P'$, the continuous dimension of $\h\H$ is $d\sim G_N^{1/2}$.

 \bibliographystyle{unsrt}

\end{document}